\documentclass[10pt,journal,twoside]{IEEEtran}
\normalsize
\usepackage{multirow}
\usepackage{amssymb}
\usepackage[dvips]{graphicx}
\usepackage{amsmath}
\usepackage{amsthm}
\usepackage{latexsym,bm}
\usepackage{color}
\usepackage{subfigure}
\usepackage{longtable}
\usepackage{accents}
\usepackage{cite}
\usepackage{enumerate}
\usepackage{arydshln}

\usepackage[ruled]{algorithm}
\usepackage{algorithmic}

\usepackage{diagbox}
\makeatletter
\def\widebar{\accentset{{\cc@style\underline{\mskip10mu}}}}
\def\Widebar{\accentset{{\cc@style\underline{\mskip8mu}}}}
\makeatother

\allowdisplaybreaks
\theoremstyle{plain}

\theoremstyle{definition}
\theoremstyle{definition} 

\begin{document}

\title{
RIS-Enabled Anti-Interference in LoRa Systems\centering
\thanks{Z. Liang, G. Cai are with the School of Information Engineering, Guangdong University of Technology, China (e-mail: Liangzk98@163.com, caiguofa2006@gdut.edu.cn).}
\thanks{J. He is with the Technology Innovation Institute, Abu Dhabi, United Arab Emirates, and also with Centre for Wireless Communications, University of Oulu, Oulu 90014, Finland (e-mail: jiguang.he@tii.ae).}
\thanks{G. Kaddoum is with the Department of Electrical Engineering, University of Qu$\acute{\mathrm{e}}$bec, Qu$\acute{\mathrm{e}}$bec, QC G1K 9H7, Canada, with the LaCIME Laboratory, $\acute{\mathrm{E}}$cole de Technologie Sup$\acute{\mathrm{e}}$rieure ($\acute{\mathrm{E}}$TS), Montreal, QC H3C 1K3, Canada, and also with the Cyber Security Systems and Applied AI Research Center, Lebanese American University, Beirut, Lebanon(e-mail: georges.kaddoum@etsmtl.ca).}
\thanks{C. Huang is with College of Information Science and Electronic Engineering, Zhejiang University, Hangzhou 310027, China (e-mail: chongwenhuang@zju.edu.cn).}
\thanks{M. Debbah is with Khalifa University of Science and Technology, P O Box 127788, Abu Dhabi, UAE (email: merouane.debbah@ku.ac.ae) and also with CentraleSupelec, University Paris-Saclay, 91192 Gif-sur-Yvette, France.}
}
\author{Zhaokun Liang,
Guofa Cai, {\em Senior Member, IEEE}, Jiguang He, {\em Senior Member, IEEE}, \\Georges Kaddoum, {\em Senior Member, IEEE}, Chongwen Huang, {\em Member, IEEE}, \\and M$\acute{\mathrm{e}}$rouane Debbah, {\em Fellow, IEEE}}

\maketitle
\begin{abstract}
It has been proved that a long-range (LoRa) system can achieve long-distance and low-power. However, the performance of LoRa systems can be severely degraded by fading. In addition, LoRa technology typically adopts an ALOHA-based access mechanism, which inevitably produces interfering signals for the target user.
To overcome the effects of fading and interference, we introduce a reconfigurable intelligent surface (RIS) to LoRa systems. In this context, both non-coherent and coherent detections are considered and their bit error rate (BER) performance analyses are conducted.
Moreover, we derive the closed-form BER expressions for the proposed system over Nakagami-$m$ fading channels.
Simulation results are used to verify the accuracy of our proposed analytical results. It is shown that the proposed system outperforms the RIS-free LoRa system, and the RIS-aided LoRa system adopting blind transmission.
Furthermore, the impacts of the spreading factor (SF), the number of reflecting elements, and the Nakagami-$m$ fading parameters are investigated.
It is shown that increasing the number of reflecting elements can remarkably enhance the BER performance, which is an affective measure for the proposed system to balance the trade-off between data rate and coverage range. We further observe that the BER performance of the proposed system is more sensitive to the fading parameter $m$ at high signal-to-noise ratios.

\end{abstract}

\section{Introduction} \label{sect:Introduction}
With the rapid development of Internet of Things (IoT), new application scenarios, such as satellite communications, unmanned aerial vehicles, industrial automation, and healthcare \cite{9509294}, have emerged.
Meanwhile, IoT technologies need to satisfy stringent requirements in terms of power consumption, quality of service (QoS), and transmission distances \cite{9424689}.
Low-power wide-area network (LPWAN), which offers the long-range connectivity with low-power consumption and operation \& maintenance cost, is designed to support various IoT deployments.
As a leading technology of LPWAN, long-range (LoRa) technology, which enables low-power, low-rate, and long-range communications, has been extensively studied, boosting tremendous commercial growth \cite{9770184}.

LoRa, which is based on chirp spread spectrum (CSS) modulation \cite{8723130}, enables the transmitted symbols to be frequency chirps that spread the whole allocated bandwidth.
The gain of LoRa depends on the spreading factor (SF), where an appropriate adjustment of the SF can lead to an optimal trade-off between data rate and transmission range. Specifically, a relatively high SF improves the transmission range at the cost of a degradation in the data rate.
In LoRaWAN, for the medium access control (MAC) layer, LoRa typically adopts an ALOHA-based mechanism \cite{9018210}. The LoRa transmitter makes use of captured packets sent sporadically at different time intervals, thus causing interference due to the packet collisions \cite{9425135}.
The interference in LoRa systems can be typically classified into two different categories: common spreading factor (co-SF) interference, among end-devices with the same SF, and inter-SF interference, among the end-devices with different SFs \cite{8480649}. Due to the spread spectrum characteristic of LoRa, these two types of interference are unavoidable. This is expected to yield a devastating impact on the LoRaWAN coverage and throughput performance as the number of LoRa devices and traffic load increases, which is the case for future IoT applications \cite{7803607}. Moreover, under network congestion, assigning a higher SF to a remote user does not improve its link performance since it is more prone to collisions \cite{8267219}.
Fortunately, thanks to the capture effect of LoRa \cite{8385552}, the packets can be successfully received even in the presence of co-SF and inter-SF interferences as long as the signal-to-interference-plus-noise ratio (SINR) exceeds the corresponding threshold. A target signal with co-SF  interference can be successfully  decoded if its received power is at least $6$ dB higher than the interfering signals \cite{8581011}. In \cite{8581011} and \cite{8903531}, theoretical bit-error-rate (BER) expressions for LoRa systems with co-SF interference were derived over the additive white Gaussian noise (AWGN) channel when the offset is constrained to integer and non-integer chip durations, respectively.

Several works were considered to enhance the performance of LoRa systems with interference \cite{9352969,9207749,9810018}.
In \cite{9352969}, coherent detection under co-SF interference was proposed to achieve a $2.5$ dB gain compared to the non-coherent detection in \cite{8903531}. Interleaved chirp spreading LoRa (ICS-LoRa) modulation was adopted to increase the capacity in \cite{9207749}. In \cite{9810018}, the authors presented a multiuser detector based on the maximum-likelihood standard, which can demodulate symbols from two interfering LoRa users. In addition, the authors designed a synchronization algorithm against co-SF interference and verified its feasibility in the GNU Radio platform. Although previous works have made significant contributions to LoRa systems, few effective solutions were proposed to resist the interference for LoRa
systems while considering fading channels.

Recently, reconfigurable intelligent surfaces (RISs) have received significant attention because of their ability to control the wireless propagation environments \cite{10006744,9140329,9410457,9500188}. A RIS consists of a large number of passive reflective elements programmed and controlled by a RIS controller, enabling beam steering toward any desired direction \cite{9110869}. In addition, the space-feed mechanism of RISs avoids excessive power loss caused by large feed networks of phased arrays. As a result, the deployment of RIS requires only very low power consumption and hardware cost \cite{9140329,9771447}.
It is worth noting that RISs are widely exploited in wireless networks to enhance system performance in terms of spectral efficiency \cite{9203956}, energy efficiency \cite{8741198}, BER \cite{9535453}, and localization accuracy \cite{9215972}. In particular, RISs were applied in wireless communication scenarios with interference to improve system performance \cite{9347448}, and suppress interference \cite{9598913}. Therefore, combining LoRa technology with RISs is a promising solution for future long-range communications, which enables a LoRa system adopting low SF with a relatively high data rate to achieve longer communication ranges
 by controlling the wireless propagation environments. Recently, RIS-aided LoRa systems with non-coherent detection were proposed to improve the BER and throughput \cite{9772404}. However, in this work, the direct link was ignored due to blockage, which is a restrictive assumption. Moreover, the authors adopted blind transmission to enhance target signal, and thus did not consider the configuration of the phase shift of the RIS. In addition, the literature only considered a single user comunicating with the gateway, which is not representative of current LoRa networks. In this context, there are multiple users transmitting simultaneously due to the ALOHA-based mechanism, resulting in severe interference, which is non-negligible to the QoS of LoRa systems.

Given the aforementioned motivations, in this paper a RIS-aided LoRa system is developed to improve system performance against the co-SF interference.
Moreover, in contrast to existing schemes in \cite{8581011,8903531,9352969,9810018}, which are limited to AWGN channels, this paper focuses on the performance analysis of the proposed system over Nakagami-$m$ fading channels, which are more realistic for IoT applications. The key contributions of this paper are summarized as follows.
\begin{itemize}
\item
We put forward a RIS-aided LoRa system to suppress co-SF interference. In the proposed system, we take advantage of the phase shift configuration ability of RISs to overcome performance degradation caused by fading and interference. Furthermore, both non-coherent and coherent detections are considered in our analytical framework.\footnote{Since coherent detection can improve the resistance of the LoRa system against co-SF interference compared to the conventional non-coherent detection \cite{9352969}, coherent detection is also considered in this paper.}
\item
We derive the closed-form BER for the proposed system with both non-coherent and coherent detections over Nakagami-$m$ fading channels. Simulation results are presented to confirm the validity of the analytical results.
Moreover, we compare the proposed system to RIS-free LoRa systems, and RIS-aided LoRa systems adopting blind transmission. Results show that the proposed system yields a performance gain compared to the above systems.
\item
We study the impact of the SF, the number of reflecting elements, and fading parameter on the BER performance of the proposed system.
It is shown that the proposed system adopting the low SF can achieve better BER performance by increasing the number of reflecting elements, which differs from the RIS-free LoRa system that enhances the BER performance by using high SF at the expense of data rate.
Furthermore, it is shown that the fading parameter $m$ is an important factor that affects the BER performance of the proposed system, especially when the target user and the co-SF interfering user use the same RIS to transmit information at high SNRs.


\end{itemize}

The rest of this paper is organized as follows. Section II describes the  foundational principle of
 LoRa modulation. Furthermore, we introduce the LoRa system with co-SF interference, followed by different types of demodulation schemes. Section III introduces the proposed system model. Section IV analyzes the performance of the proposed system. Section V presents the simulation results and discussions. Finally, Section VI concludes this paper.

$\mathit{Notation}$: $j=\sqrt{-1}$.  $|\cdot|$ denotes absolute value operation. The probability of an event is denoted by $\mathrm{Pr\left(  \cdot  \right)} $. ${f_X}( \cdot )$ is the probability density function (PDF) of the random variable (RV) $X$. $\mathrm{E}(X)$ and $\mathrm{V}(X)$ are the mean and variance of the RV $X$, respectively, and ${\mu _X}\left( \iota \right) = \mathrm{E}\left( {{X^\iota}} \right)$ is the $\iota$-th moment of RV $X$. $\mathcal{CN}(\mu,\sigma  ^2)$ and $\mathcal{N}(\mu,\sigma  ^2)$ denote the complex Gaussian distribution and Gaussian distribution with mean $\mu$ and variance $\sigma  ^2$, respectively. $*$ denotes the complex conjugate operation. ${G_{{x_1}}}\left(  \cdot  \right)$ is the modified Bessel function of the second kind. ${D_{{x_2}}}\left(  \cdot  \right)$ denotes the parabolic cylinder function. $\gamma(\cdot)$ is the Gamma function. $Q\left( \cdot \right)$ is the Gaussian $Q$-function. $\Re(x)$ denotes the real part of complex number $x$. The PDF of the Rayleigh and Rice distributions are denoted by ${f_{{\rm{Ra}}}}\left( { \cdot ;\sigma } \right)$ and ${f_{{\rm{Ri}}}}\left( { \cdot ;\sigma ,\mathchar'26\mkern-10mu\lambda } \right)$, respectively, where $\sigma $ is scale parameter and $\mathchar'26\mkern-10mu\lambda $ is the Rice factor.
\section{The LoRa Modulation} \label{sect:lora}
In this section, we briefly recall the foundational principles of conventional LoRa systems. Furthermore, we introduce the LoRa system with co-SF interference, followed by different types of demodulation schemes.

\subsection{LoRa Basic Modulation} \label{sect:lora_modulation}
LoRa adopts CSS-based modulation. In discrete-time baseband signals, one LoRa sample lasts $T_{s}=\frac{1}{B}$, where $B$ denotes the channels bandwidth.\footnote{The bandwidth $B$ of LPWAN is usually set to $125$ kHz, $250$ kHz, or $500$ kHz depending on the practical application.} One LoRa symbol consists of $K=2^{SF}$ samples and the duration of each LoRa symbol is $T_{c}=KT_{s}$, where $SF \in \{7,...,12\}$ denotes the spreading factor. The LoRa encoder maps a decimal symbol $c \in \mathcal{C} $ to $SF$ bits, where $\mathcal{C} = \{0,1,...,K-1\}$. 
The frequency of the transmitted symbol $c$ increases linearly with the slope $l = \frac{{{B^2}}}{K}$ from the initial frequency $f=\frac{B}{K}c$ to $  \frac{B}{2}$, and then drops to $ - \frac{B}{2}$, while in the remaining symbol duration, the frequency continues to change linearly with the slope  $l$ from $ - \frac{B}{2}$ to $f$. Thus, the discrete-time baseband signal is given by \cite{8723130}
\begin{align}
\label{eq:signal}
{x_c}[n] \!=\! \sqrt {\frac{{{1}}}{{{K}}}} \exp \left[ {j2\pi \left( {\frac{{{n^2}}}{{2K}} + \left( {\frac{c}{K} - \frac{1}{2}} \right)n} \right)} \right],n \in {\mathcal C},
\end{align}
where $n=0,1,2,...,K-1$ denotes the sample index.

At the gateway, the received signal is given by
\begin{align}
\label{eq:signal_rc}
r_c[n]= h_c\sqrt {{E_c}} x_c[n] + z[n],n \in {\mathcal C},
\end{align}
where $h_c$ is the fading coefficient from the LoRa transmitter to the gateway, $E_c$ denotes the energy of the transmitted symbol, and $z[n] \sim \mathcal{CN}(0,N_0)$ is the AWGN.


To recover the signal, the received signal is multiplied by the complex conjugate of the basic chirp $x_0^*[n]$, which is also called dechirping, and then a discrete Fourier transform (DFT) is performed to the dechirped signal. The DFT output is given by
\begin{align}
\label{eq:DFToutput1}
{{\rm {O}}_i}& = {\rm{DFT}}\left( {{x_s}[k]x_0^*[k]} \right)\nonumber \\
& = \frac{{\left| {{h_c}} \right|\sqrt {{E_c}}  }}{K}\sum\limits_{n = 0}^{K - 1} {\exp \left( {j2\pi n\frac{{\left( {c - i} \right)}}{K} + j{\varphi _c}} \right)}  + Z[i]\nonumber\\
& = \left\{ {\begin{array}{*{20}{c}}
{\left| {{h_c}} \right|\sqrt {{E_c}}  \exp \left( {j{{\varphi _c}}} \right) + Z[i],}&{i = c,}\\\
{\!\!\!\!\!\!\!\!\!\!\!\!\!\!\!\!\!\!\!\!\!\!\!\!\!\!\!\!\!\!\!\!\!\!\!\!\!\!\!
\!\!\!\!\!\!\!\!\!\!\!\!\!\!\!\!\!\!\!\ Z[i],}&{\!\!\ i \ne c,}
\end{array}} \right.
\end{align}
where $\left| {{h_c}} \right|$ and ${\varphi _c}$ are the magnitude and phase of the channel coefficient $h_c$, respectively, and $Z[i] \sim \mathcal{CN}(0,N_0)$, $i \in \{0,1,2,...,K-1\}$ is the possible symbol.


Non-coherent detection is to find the index $\hat{c}$ with the largest magnitude of ${\rm {O}}_i$ \cite{8903531}, obtained as
\begin{align}
\label{eq:de_max_2}
\hat c = \arg \mathop {\max }\limits_{i \in \mathcal{C}} \left( {\left| {{{\rm{O}}_i}} \right|} \right).
\end{align}
Coherent detection requires compensation for the channel phase shift. Since the channel phase shift ${\varphi _c}$ can be estimated using the preamble of LoRa symbols\cite{9352969,8835951}, it is feasible to obtain the compensation phase shift $\tilde \varphi  =  - {\varphi _c}$. Thus, the coherent receiver finds the index $\hat{c}$ with the maximum real part value of ${{\rm{O}}_i}\exp \left( {j\tilde \varphi } \right)$ \cite{9352969}, given by
\begin{align}
\label{eq:de_max_3}
\hat c = \arg \mathop {\max }\limits_{i \in {\cal C}} \Re \left( {{{\rm{O}}_i}\exp \left( {j\tilde \varphi } \right)} \right).
\end{align}
\subsection{LoRa Systems with co-SF Interference} \label{sect:interference_rayleigh}
\begin{figure}[t]
\center
\includegraphics[width=3in,height=1.75in]{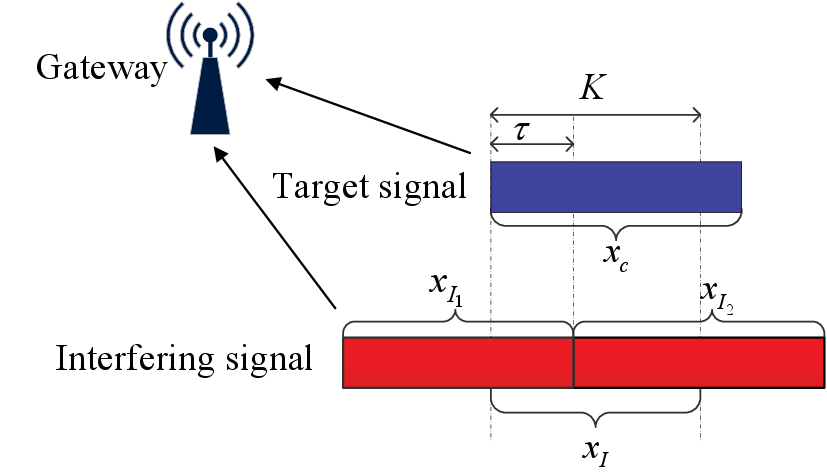}
\caption{Block diagram of interfering signal model.}
\label{fig_int_system_model}
\vspace{-0.45cm}
\end{figure}
We consider an IoT application with a large number of devices adopting the ALOHA-based mechanism to access channels in LoRaWAN. In this context, a gateway attempts to recover the symbol of a target user in the presence of an interfering LoRa user, as shown in Fig.~\ref{fig_int_system_model}.
As mentioned earlier, the interference in LoRa systems are classified into inter-SF interference and co-SF interference. It is noted that although the effect of inter-SF interference has been proven to be non-negligible, especially in the case of higher SF, it is very different from the co-SF interference because of the spreading gain \cite{8886735,9352969}. The results in \cite{8886735} show that due to the orthogonality of LoRa signals with different SFs, the dechirped signals with inter-SF interference exhibit low spectral density in a broadband spectrum in the frequency domain. Thus, inter-SF interference can be considered as white noise.
In this paper, we consider the scenario where the target user incurs co-SF interference, which yields the most serious impact on the target signal. For simplicity,
we consider one co-SF interfering user since when there are multiple co-SF interfering users, the strongest interfering user generally has the most devastating impact on the BER.

Based on some low-complexity synchronization methods for LoRa systems presented in \cite{9352969}, we assume the LoRa gateway is perfectly synchronized with the target user, and the interfering user is neither synchronized with the target user nor the gateway.
Thus, the interfering signal is generally composed of two different LoRa symbols $x_{I_1}$ and $x_{I_2}$, $I_1$, $I_2 \in \mathcal{C}$ due to the lack of perfect time synchronization, as depicted in Fig.~\ref{fig_int_system_model}. There exists a time difference between the interferer and the target user. Let's define a random offset $\tau$ between the target signal $x_c$ and the interfering signal $x_{I_2}$. Namely, the time difference between the first frequency bin of the signal $x_{I_2}$  and the first frequency bin of the signal $x_c$  is $\tau$.
To simplify the analysis, we assume that $\tau$ is an integer, following a uniform distribution in $\mathcal{C}$. Therefore, referring to  (\ref{eq:signal}), the interfering signal is given by
\begin{equation}
\begin{aligned}
\label{eq:inter_signal}
x{}_I[n] = \left\{ {\begin{array}{*{20}{c}}
  {x_{{I_1}}}[n],&{0 \leqslant n < \tau ,} \\
  {x_{{I_2}}}[n],&{\ \tau  \leqslant n < K}.
\end{array}} \right.
\end{aligned}
\end{equation}

The received signal can thus be expressed as
\begin{equation}
\begin{aligned}
\label{eq:recev_signal}
y[n]=h_c\sqrt {{E_c}}x_c[n]+h_I\sqrt {{E_I}} x_I[n]+z[n],
\end{aligned}
\end{equation}
where $h_I$ is the fading coefficient of the channel from the interfering user to the gateway and $E_I$ denotes the transmitted energy of the interference LoRa symbol. We assume that the target signal and the interfering signal have the same transmitted energy, i.e., ${E_c} = {E_I} = 1$.

Without loss of generality, we assume that the number of samples of $x_{I_2}[n]$ exceeds that of $x_{I_1}[n]$, i.e., $0\leqslant \tau \leqslant \frac{K}{2}$.
According to (\ref{eq:DFToutput1}), the non-coherent demodulation metric can be obtained as
\begin{align}
\label{eq:correlator_interference}
\!\!\left|\! {{{\dot \Upsilon }_i}} \right| \!=\! \left\{ {\begin{array}{*{20}{c}}
{\!\!\left| {  \left| {{h_c}} \right|\!\sqrt {\!{E_c}}  \exp \left( {j{\varphi _c}} \right)\! + \!\left| {{h_I}} \right|\!\sqrt {\!{E_I}} {\Psi _i}\exp\! \left( {j{\varphi _I}} \right) \!+\! Z[i]} \right|\!,}&{\!\!\!i = c,}\\
{\!\!\!\!\!\!\!\!\!\!\!\!\!\!\!\!\!\!\!\!\!\!\!\!\!\!\!\!\!\!\!\!\!\!\!\!
\!\!\!\!\!\!\!\!\!\!\left| { \left| {{h_I}} \right|\sqrt {{E_I}}  {\Psi _i}\exp \left( {j{\varphi _I}} \right) + Z[i]} \right|,}&{\!\!\!\!\!\ i \ne c,}
\end{array}} \right.
\end{align}
where $i \in \mathcal{C}$, $\left| {{h_I}} \right|$ and ${\varphi _I}$ are the magnitude and phase shift of the fading coefficient $h_I$, respectively, and ${\Psi _i}$ is the cross-correlation interference term generated by (\ref{eq:inter_signal}).

The coherent receiver is required to compensate for the channel phase shift and then get its real part value. Thus, the demodulation metric for coherent detection is given by
\begin{align}
\label{eq:interference_coherent_RIS_free}
\!\!\Re\! \left( \!{{{\dot \Upsilon }_i}\exp \left( {j\tilde \varphi } \right)} \right)\! =\! \left\{ {\begin{array}{*{20}{c}}
{\!\!\left| {{h_c}} \right|\sqrt {{E_c}}  + \left| {{h_I}} \right|\sqrt {{E_I}} {\Psi _i}\cos \vartheta  + \eta [i],}&\!\!\!{i = c,}\\
{\!\!\!\!\!\!\!\!\!\!\!\!\!\!\!\!\!\!\!\!\!\!\!\!\!\!\!\!\!\!\!\!\left| {{h_I}} \right|\sqrt {{E_I}} {\Psi _i}\cos \vartheta  + \eta [i],}&\!\!\!{i \ne c,}
\end{array}} \right.
\end{align}
where $\vartheta  = {\varphi _I} + \tilde \varphi $, and $\eta [i] \sim {\cal N}(0,\frac{{{N_0}}}{2})$.

To evaluate the impact of the interfering signal on the BER of the LoRa system, it is  essential to evaluate the magnitudes $\left| {{\Psi _i}} \right|$. Combining (\ref{eq:DFToutput1}) and (\ref{eq:inter_signal}), ${\Psi _i}$ can be expressed as
\begin{align}
\label{eq:interference_term}
{\Psi _i} =  \left. {{{ {\rm O}}_i}} \right|_0^{\tau  - 1}  + \left. {{{ {\rm O}}_i}} \right|_\tau ^{K - 1}  ,
\end{align}
where
\begin{align}
&\left. {{{{\rm O}}_i}} \right|_0^{\tau  - 1} = \frac{1}{{K}}\sum\limits_{l  = 0}^{\tau  - 1} {\exp \left( {\frac{{j2\pi l \left( {i - {I_1}} \right)}}{{K}}} \right)}   ,\nonumber \\
&\left. {{{{\rm O}}_i}} \right|_\tau ^{K - 1}  = \frac{1}{{K}}\sum\limits_{l  = \tau }^{K - 1} {\exp \left( {\frac{{j2\pi l \left( {i - {I_2}} \right)}}{{K}}} \right)}     \label{eq:interference_term_2}.
\end{align}

Following the geometric series summation rule, (\ref{eq:interference_term}) can be rewritten as
\begin{align}
\label{eq:sum_interference_term}
{\Psi _i} = {\rm{ }}\frac{1}{K}\left( {\frac{{\sin \left( {\pi \frac{{i - {I_1}}}{K}\tau } \right)}}{{\sin \left( {\pi \frac{{i - {I_1}}}{K}} \right)}} + \frac{{\sin \left( {\pi \frac{{i - {I_2}}}{K}\left( {K - \tau } \right)} \right)}}{{\sin \left( {\pi \frac{{i - {I_2}}}{K}} \right)}}} \right).
\end{align}

Exploiting the absolute value inequality, an upper bound ${\chi _i}$ can be obtained as
\begin{align}
\label{eq:bound}
\left| {{\Psi _i}} \right| \le {\chi _i} \!=\! \frac{1}{K} \times \!\left( {\underbrace {\left| {\frac{{\sin \left( {\pi \frac{{i - {I_1}}}{K}\tau } \right)}}{{\sin \left( {\pi \frac{{i - {I_1}}}{K}} \right)}}} \right|}_{{\Delta _1}} + \underbrace {\left| {\frac{{\sin \left( {\pi \frac{{i - {I_2}}}{K}\left( {K - \tau } \right)} \right)}}{{\sin \left( {\pi \frac{{i - {I_2}}}{K}} \right)}}} \right|}_{{\Delta _2}}} \right),
\end{align}
where ${\chi  _i}$ exhibits a peak cross-correlation interference at a certain  bin $\tilde i = {\arg _{i \in {\cal C}}}\max \left( {{\chi _i}} \right)$ for each realization of $I_1$, $I_2$, and  $\tau$ \cite{8581011}. Thus, the cross-correlation interference over the remaining bins $i\ne \tilde i$ is negligible, ${\chi  _i}$ is can be approximated as
\begin{align}
\label{eq:bound_app_1}
{\chi _i} \approx \left\{ {\begin{array}{*{20}{c}}
  {{\chi _{\tilde i}},}&{i = \tilde i} \\
  {0,}&{i \ne \tilde i}.
\end{array}} \right.
\end{align}

It is noted that the expression  ${{\sin (\rho x)} \mathord{\left/
 {\vphantom {{\sin (\rho x)} {\sin (x)}}} \right.
 \kern-\nulldelimiterspace} {\sin (x)}}$ is equal to its maximum  $\rho $ at $x = 0$. Similarly, ${{\Delta _1}}$ and ${{\Delta _2}}$ have maximum values $\tau / K$ and $(K-\tau)/K$ at $i=I_1$ and $i={I_2}$, respectively. Since we assume $0\leqslant \tau \leqslant K/2$, for (\ref{eq:bound}), peak cross-correlation interference is  more likely to occur at ${\tilde i}={I_2}$. Hence, (\ref{eq:bound_app_1}) can be further approximated as
\begin{align}
\label{eq:bound_app_2}
{\chi _{\tilde i}} \approx {\chi _{{I_2}}} = \frac{1}{K}\left( {\left| {\frac{{\sin \left( {\pi \frac{{{I_2} - {I_1}}}{K}\tau } \right)}}{{\sin \left( {\pi \frac{{{I_2} - {I_1}}}{K}} \right)}}} \right| + K - \tau } \right).
\end{align}

We observe that $\chi  _{I_2}$ is a function of ${I_1}$, ${I_2}$, and $\tau$ and averaging the cumulative sum of ${I_1}$, ${I_2}$, and $\tau$ incurs a high  complexity. Let $I={I_2}-{I_1}$, (\ref{eq:bound_app_2}) can be further written as
\begin{align}
\label{eq:bound_app_I}
{\chi _{{I}}}  =\chi _{{I_2}}  \approx\frac{1}{K}\left( {\left| {\frac{{\sin \left( {\pi \frac{I}{K}\tau } \right)}}{{\sin \left( {\pi \frac{I}{K}} \right)}}} \right| + K - \tau } \right).
\end{align}

According to (\ref{eq:bound_app_1}) and (\ref{eq:bound_app_I}), (\ref{eq:correlator_interference}) can be approximated as
\begin{align}
\label{eq:RIS_free_inter_non_budengyu}
&\left| {{{\dot \Upsilon }_{i|c \ne {I_2}}}} \right| \approx \left\{ {\begin{array}{*{20}{l}}
{\left| {\left| {{h_I}} \right|\sqrt {{E_I}}  {\chi _I}\exp \left( {j{\varphi _I}} \right) + Z[i]} \right|,}&{i = {I_2,}}\\
{\left| {\left| {{h_c}} \right|\sqrt {{E_c}}  \exp \left( {j{\varphi _c}} \right) + Z[i]} \right|,}&{i = c,}\\
{\left| {Z[i]} \right|,}&{i \ne {I_2},c.}
\end{array}} \right.
\\&\label{eq:RIS_free_inter_non_dengyu}\left| {{{\dot \Upsilon }_{i|c = {I_2}}}} \right| \approx \left\{ {\begin{array}{*{20}{c}}
{\begin{array}{*{20}{l}}
{\left| {\left| {{h_c}} \right|\sqrt {{E_c}} \exp \left( {j{\varphi _c}} \right) + \left| {{h_I}} \right|\sqrt {{E_I}} {\chi _I}\exp \left( {j{\varphi _I}} \right)} \right.}\\
{\left. { + Z[i]} \right|\!,\ \ \ \!\!i = {I_2},}
\end{array}}\\
{\!\!\!\!\!\!\!\!\!\!\!\!\!\!\!\!\!\!\!\!\!\!\!\!\!\!\!\!\!\!\!\!\!\!\!\!\!\!\!\!
\!\!\!\!\!\!\!\!\!\!\!\!\!\!\!\!\!\!\!\!\!\!\!\!\!\!\!\!\!\left| {Z[i]} \right|,\ \ \ \!i \ne {I_2}.}
\end{array}} \right.
\end{align}
where (\ref{eq:RIS_free_inter_non_budengyu}) and (\ref{eq:RIS_free_inter_non_dengyu}) differentiate the two cases when the peak cross-correlation interference and the target symbol $c$ appear at different bin outputs, i.e., $c \neq I_2$,
and when the symbol and peak interference appear at the same bin outputs, i.e., $c = I_2$, respectively.
Non-coherent receiver selects the bin index with the largest of $\left| {{{\dot \Upsilon }_i}} \right|$, obtained as
\begin{align}
\hat c = \mathop {\arg \max }\limits_{i \in C} \left( {\left| {{{\dot \Upsilon }_i}} \right|} \right).
\end{align}

The coherent demodulation metric from (\ref{eq:interference_coherent_RIS_free}) can be further approximated as follows:
\begin{align}
&\label{eq:RIS_free_inter_co_budengyu}
\!\!\Re \left( {{{\dot \Upsilon }_{i|c \ne {I_2}}}\exp \left( {j\tilde \varphi } \right)} \right) \approx \left\{ {\begin{array}{*{20}{l}}
{\left| {{h_I}} \right|\sqrt {{E_I}}  {\chi _I}\cos \vartheta  + \eta [i],}{\ \!i = {I_2,}}\\
{\left| {{h_c}} \right|\sqrt {{E_c}}  \cos \vartheta  + \eta [i],}{\ \ \ \ i = c,}\\
{   \ \!\eta [i],}{\ \ \ \ \ \ \ \ \ \ \ \ \ \ \ \ \ \ \ \ \ \ \ \  i \ne {I_2},c.}
\end{array}} \right.
\\&\label{eq:RIS_free_inter_co_dengyu}\Re \left( {{{\dot \Upsilon }_{i|c = {I_2}}}\exp \left( {j\tilde \varphi } \right)} \right) \approx \left\{ {\begin{array}{*{20}{c}}
\begin{array}{l}
\left| {{h_c}} \right|\sqrt {{E_c}}  + \left| {{h_I}} \right|\sqrt {{E_I}} {\chi _I}\cos \vartheta \\
 + \eta [i],\ \!i = {I_2},
\end{array}\\
{\!\!\!\!\!\!\!\!\!\!\!\!\!\!\!\!\!\!\!\!\!\!\!\!\!\!\!\!\!\!\!\!\!\!\!\!\!\!\!\
\eta [i],\ \ i \ne {I_2}.}
\end{array}} \right.
\end{align}
Coherent receiver selects the bin index with the maximum of $\Re \left( {{{\dot \Upsilon }_{i}}\exp \left( {j\tilde \varphi } \right)} \right)$, expressed as
\begin{align}
\hat c & = \arg \mathop {\max }\limits_{i \in {\cal C}} \Re \left( {{{\dot \Upsilon }_{i}}\exp \left( {j\tilde \varphi } \right)} \right).
\end{align}

Given that $SF \ge 7$ and $K \gg 1$, we assume that the cross-correlation interference appears on a different bin than the target signal. Accordingly, the case of $c \ne {I_2}$ is considered to analyze the BER of the proposed system.
\section{Proposed RIS-aided LoRa System} \label{sect:system model}
\begin{figure}
\centering
\subfigure[]{
\includegraphics[width=3.2in,height=1.9in]{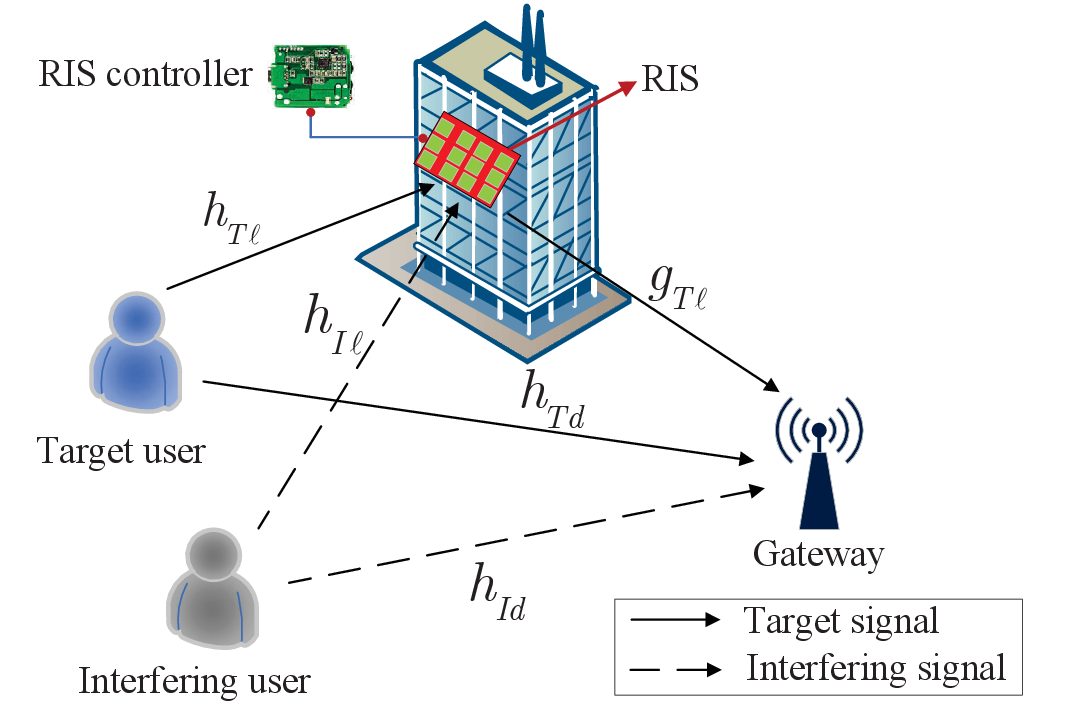}
}
\subfigure[]{
\includegraphics[width=3.6in,height=1.85in]{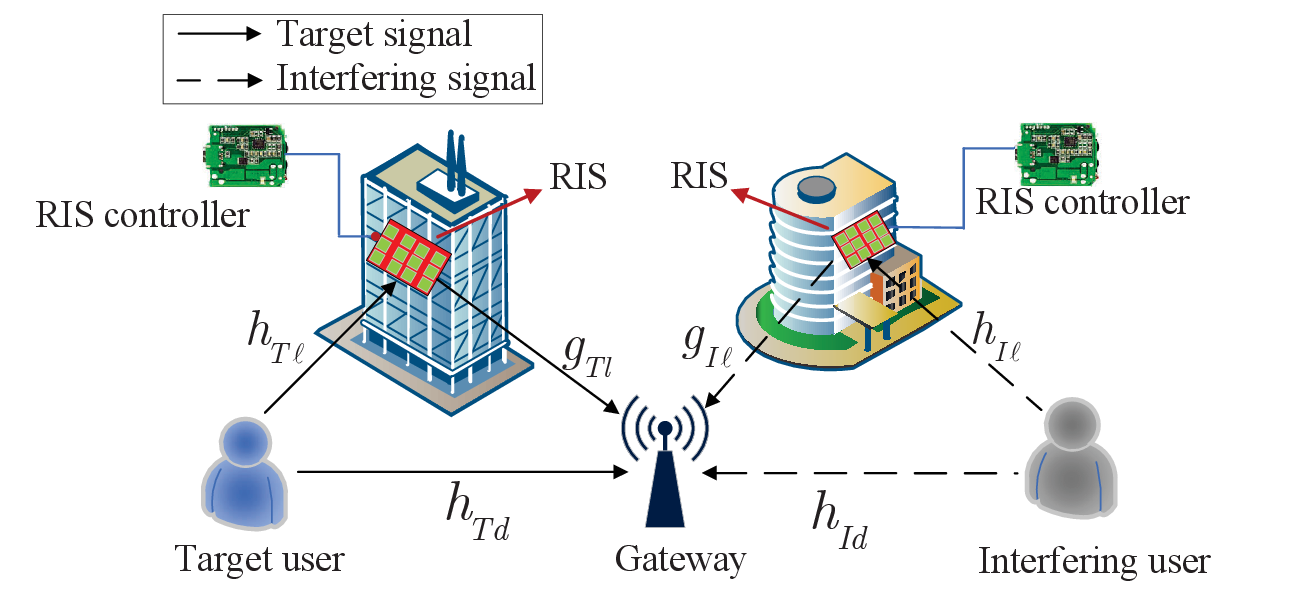}
}
\caption{{Block diagram of the proposed RIS-aided LoRa uplink system, where (a) Case A:  the target user and the interfering user adopt the same RIS and (b) Case B: the target user and the interfering user adopt different RISs.}}
\label{fig:fig_RIS_int_system_model}
\end{figure}

In this section, we consider a RIS-aided uplink LoRa system in which a LoRa gateway serves as a target user in the presence of an interfering LoRa user, as shown in Fig.~{\ref{fig:fig_RIS_int_system_model}}. Both users communicate with the LoRa gateway through a direct link and reflection links. Each RIS has $N$ passive reflective elements, and is programmed by a controller to reconfigure the phase shifts via a backhaul link.\footnote{It should be noted that backhaul links typically have very high transmission rates to achieve the real-time configuration of RIS phase shifts \cite{9138463,9558795}.} It is noted that RISs bring not only stronger target signals, but also more severe interfering signals. Therefore, we consider two cases. Case A: The target user and the interfering user adopt the same RIS to deliver information, shown in Fig.~\ref{fig:fig_RIS_int_system_model} (a); Case B: The target user and the interfering user adopt different RISs to transmit information, as shown in Fig.~\ref{fig:fig_RIS_int_system_model} (b).


In Fig.~{\ref{fig:fig_RIS_int_system_model}} (a), the RIS is paired with the target user to enhance the desired signals, but more severe interfering signals are inevitably introduced. Let ${h_{T\ell }}$, ${g_{T\ell }}$, and ${h_{Td}}$ denote fading coefficients from the target user to the $\ell$-th reflecting element of the RIS, from that reflecting element to the LoRa gateway, and from the target user to the LoRa gateway, respectively. Let ${h_{I\ell }}$ and ${h_{Id}}$ denote the fading coefficients from the interfering user to the $\ell$-th RIS' reflecting element and from the interfering user to the LoRa gateway, respectively.

In Fig.~{\ref{fig:fig_RIS_int_system_model}} (b), we consider that both users and RISs are pairwise. In such case, the RIS boosts the signals of its paired users. We follow the previous notation and add ${g_I}_\ell $ as the fading coefficient from the $\ell$-th reflecting element of the RIS to the LoRa gateway.

Although there are several works  on the BER of LoRa systems in the presence of interference \cite{8581011,8903531,9352969,9810018}, they consider AWGN channels, which is unrealistic. To this end, the Nakagami-$m$ fading channel is adopted in this paper.\footnote{Nakagami-$m$ fading channels can comprise both  line-of-sight (LOS) and non-line-of-sight (NLOS) propagation environments by changing the parameter $m$, which can be easily fitted into different channel conditions. Thus, it is very suitable for IoT applications \cite{9406837,8077766}.}
To facilitate the analysis, we define the channels as follows:
\begin{align}
{h_{Td}} &= \left| {{h_{Td}}} \right|\exp \left( {j{\theta _{{h_{Td}}}}} \right),\label{eq:T_d} \\
{h_{Id}} &= \left| {{h_{Id}}} \right|\exp \left( {j{\theta _{{h_{Id}}}}} \right),\label{eq:I_d} \\
{h_{T\ell }} &= \left| {{h_{T\ell }}} \right|\exp \left( {j{\theta _{{h_{T\ell }}}}} \right)
,\label{eq:T_h} \\
{g_{T\ell }} &= \left| {{g_{T\ell }}} \right|\exp \left( {j{\theta _{{g_{T\ell }}}}} \right)
 ,\label{eq:T_g} \\
{h_{I\ell }} &= \left| {{h_{I\ell }}} \right|\exp \left( {j{\theta _{{h_{I\ell }}}}} \right)
,\label{eq:I_h} \\
{g_{I\ell }}& = \left| {{g_I}_\ell } \right|\exp \left( {j{\theta _{{g_{I\ell }}}}} \right)
 ,\label{eq:I_g} \\
{\theta _{T\ell }}& = {\theta _{{h_{T\ell }}}} + {\theta _{{g_{T\ell }}}}       ,\label{eq:phase_T_hg} \\
{\theta _{I\ell }}& = {\theta _{{h_{I\ell }}}} + {\theta _{{g_{T\ell }}}}        ,\label{eq:phase_I1_hg} \\
{\theta _{I'\ell }} &= {\theta _{{h_{I\ell }}}} + {\theta _{{g_{_{I\ell }}}}}      ,\label{eq:phase_I2_hg}
\end{align}
where $\left| {{h_{Td}}} \right|$, $\left| {{h_{Id}}} \right|$, $\left| {{h_{T\ell }}} \right|$, $\left| {{g_{T\ell }}} \right|$, $\left| {{h_{I\ell }}} \right|$ and $\left| {{g_I}_\ell } \right|$ are the magnitudes of ${h_{Td}}$, ${h_{Id}}$, ${h_{T\ell }}$, ${g_{T\ell }}$, ${h_{I\ell }}$, and ${g_{I\ell }}$, respectively,
${{\theta _{{h_{Td}}}}}$, ${{\theta _{{h_{Id}}}}}$, $\theta_{{h} _{T\ell }}$, $\theta_{{g} _{T\ell }}$, $\theta_{{h} _{I\ell }}$, and $\theta_{{g} _{I\ell }}$ are the corresponding phases, which are uniformly distributed within $[0,2\pi)$.
We adopt the Nakagami-$m$ fading channel model in this paper, where $\left| {{h_{Td}}} \right|\sim{\rm{Nakagami}}\left( {{m_1},1} \right)$, $\left| {{h_{Id}}} \right|\sim{\rm{Nakagami}}\left( {{m_2},1} \right)$, $\left| {{h_{T\ell }}} \right|\sim{\rm{Nakagami}}\left( {{m_h}_{_{T }},1} \right)$, $\left| {{g_{T\ell }}} \right|\sim{\rm{Nakagami}}\left( {{m_g}_{_{T }},1} \right)$,
$\left| {{h_{I\ell }}} \right|\sim{\rm{Nakagami}}\left( {{m_h}_{_{I }},1} \right)$ and $\left| {{g_{I\ell }}} \right|\sim{\rm{Nakagami}}\left( {{m_g}_{_{I\ }},1} \right)$, with ${m_1}$, ${m_2}$, ${m_{{h_T}}}$, ${m_{{g_T}}}$, ${m_{{h_I}}}$, and ${m_{{g_I}}}$ being the shape parameters of the channel coefficients ${h_{Td}}$, ${h_{Id}}$, ${h_{T\ell }}$, ${g_{T\ell }}$ , ${h_{I\ell}}$, and ${g_{I\ell }}$, respectively.
\subsection{RIS-Aided Transmission for Case A} \label{sect:RIS Intelligent_A}
In such a scenario, the LoRa gateway receives a superposition of the target signal and interfering signal via the RIS. Therefore, the received signal at the gateway can be obtained as
%
\begin{align}
\label{eq:int_y1_1}
{y_A}\left[ n \right] &= \left( {\sum\limits_{\ell  = 1}^N {{h_{T\ell }}\exp (j{\omega _{T\ell }}){g_{T\ell }} + {h_{Td}}} } \right)\sqrt {{E_c}} {x_c}[n] \nonumber \\&+ \left( {\sum\limits_{\ell  = 1}^N {{h_{I\ell }}\exp (j{\omega _{T\ell }}){g_{T\ell }} + {h_{Id}}} } \right)\sqrt {{E_I}} {x_I}[n] + z[n],
\end{align}
where ${\omega _{T\ell }} \in \left[ {0,2\pi } \right)$ denotes the phase shift generated by the $\ell$-th reflecting element of the RIS paired with the target user.\footnote{In practice, the realization of the RIS phase shifts is determined by the electronics in the RIS, such as diodes and field effect transistors, leading to a finite number of discrete values possible for the phase shift. In this paper, to maximize the RIS beamforming gain, we assume that the phase shifts are continuous between $\left[ {0,2\pi } \right)$ \cite{9140329,9558795}.} We consider that the RIS has perfect knowledge of the target user's CSI, and can therefore achieve the optimal configuration of the RIS phase shifts to maximize the strength of the target signal.\footnote{Recent literatures have proved that it is feasible to perform channel estimation to obtain the CSI via deep learning and optimization techniques. Some low-complexity and high-accuracy algorithms were extensively applied in multiuser systems \cite{9627230} and mmWave MIMO systems \cite{9398559}.} As a result, we have ${\omega _{T\ell }} =  - {\theta _{T\ell }} + {\theta _{{h_{Td}}}}$  and (\ref{eq:int_y1_1}) can be rewritten as
\begin{align}
\label{eq:int_yA1}
{y_A}\left[ n \right]{\rm{ }}& = \underbrace {\left( {\sum\limits_{\ell  = 1}^N {\left| {{h_{T\ell }}} \right|\left| {{g_{T\ell }}} \right|}  + \left| {{h_{Td}}} \right|} \right)\exp \left( {j{\theta _{{h_{Td}}}}} \right)}_H\sqrt {{E_c}} {x_c}[n] \nonumber\\
&+ \underbrace {\left( {\sum\limits_{\ell  = 1}^N {{h_{I\ell }}\exp (j{\omega _{T\ell }}){g_{T\ell }}}  + {h_{Id}}} \right)}_{{H_{ A}}}\sqrt {{E_I}} {x_I}[n] + z[n].
\end{align}
In (\ref{eq:int_yA1}), we define $H = \left| H \right|\exp (j{\theta _{{h_{Td}}}})$ and ${H_A} = \left| {{H_A}} \right|\exp (j{\theta _{{H_A}}})$, where ${\theta _H}_{_{ A}}$ is uniformly distributed in $[0,2\pi)$, $\left| H \right|$ and $\left| {{H_{ A}}} \right|$ are the magnitudes of ${H}$ and ${H_{ A}}$, respectively.
\subsection{RIS-Aided Transmission for Case B} \label{sect:RIS Intelligent_B}
In contrast to Case A, here, the target and the interfering users adopt different RISs. The received signal can be expressed as
\begin{align}
\label{eq:int_yB1}
{y_B}\left[ n \right] &= \left( {\sum\limits_{\ell  = 1}^N {{h_{T\ell }}\exp (j{\omega _{T\ell }}){g_{T\ell }} + {h_{Td}}} } \right)\sqrt {{E_c}} {x_c}[n] \nonumber \\
&+ \left( {\sum\limits_{\ell  = 1}^N {{h_{I\ell }}\exp (j{\omega _{I\ell }}){g_{I\ell }} + {h_{Id}}} } \right)\sqrt {{E_I}} {x_I}[n] + z[n],
\end{align}
where ${\omega _{I\ell }} \in \left[ {0,2\pi } \right)$ is the phase shift of the $\ell$-th reflecting element of the RIS paired with the interfering user. Unlike Case A, we assume that the RISs, paired with the target user and the interfering user, have the corresponding CSIs. Thus, the RISs can provide optimal phase shifts for the target signal and the interfering signal, i.e., ${\omega _{T\ell }} =  - {\theta _{T\ell }} + {\theta _{{h_{Td}}}}$ and ${\omega _{I\ell }} =  - {\theta _{I\ell }} + {\theta _{{h_{Id}}}}$. As a result, (\ref{eq:int_yB1}) can be rewritten as
\begin{align}
\label{eq:int_yB2}
{y_B}\left[ n \right]& = \underbrace {\left( {\sum\limits_{\ell  = 1}^N {\left| {{h_{T\ell }}} \right|\left| {{g_{T\ell }}} \right|}  + \left| {{h_{Td}}} \right|} \right)\exp \left( {j{\theta _{{h_{Td}}}}} \right)}_H\sqrt {{E_c}} {x_c}[n]{\rm{ }} \nonumber\\
&+ \!\underbrace {\left( {\sum\limits_{\ell  = 1}^N {\left| {{h_{I\ell }}} \right|\left| {{g_I}_\ell } \right|}  + \left| {{h_{Id}}} \right|} \right)\exp \left( {j{\theta _{{h_{Id}}}}} \right)}_{{H_{ B}}}\sqrt {{E_I}} {x_I}[n]\! + \!z[n].
\end{align}
In (\ref{eq:int_yB2}), we define ${H_B} = \left| {{H_B}} \right|\exp (j{\theta _{{h_{Id}}}})$, where $\left| {{H_B}} \right|$ is the magnitude of ${H_B}$.

{{\em Remark~1:}
In this paper, we focus on the gains achieved by integrating RISs into LoRa systems with co-SF interference at the PHY layer. Recently, the issue of how a RIS can be integrated into the MAC layer is considered in \cite{10109667}. The authors proposed a new MAC protocol to solve the random access problem in RIS-aided systems. This protocol has a 60\% higher throughput than the traditional ALOHA protocol. Building on the literature, the question of how to integrate the RIS into the ALOHA protocol of the LoRa system is an interesting and challenging problem, which will be studied in the future.

\section{Performance Analyses} \label{sect:perfomnanceanalysis}
In this section, the BER performance of the proposed RIS-aided LoRa system is analyzed.
%
\subsection{BER Analysis for Case A} \label{sect:BER Analysis_Int}
\subsubsection{Non-coherent detection} \label{sect:BER Non-coherent detection_caseA}
According to (\ref{eq:RIS_free_inter_non_budengyu}) and (\ref{eq:int_yA1}), the demodulation metric for non-coherent detection can be obtained as
\begin{align}
\label{eq:RIS_interferenceA_magnitude}
\left| { {{{\ddot \Upsilon }_{i|c \ne {I_2}}^A}}} \right|  \approx \left\{ {\begin{array}{*{20}{l}}
{\left| {\left| {{H_A}} \right|\sqrt {{E_I}}  {\chi _I}\exp \left( {j{\theta _{{H_{ A}}}}} \right) + Z[i]} \right|,}&{i = {I_2,}}\\
{\left| {\left| H \right|\sqrt {{E_c}}  \exp \left( {j{\theta _{{h_{Td}}}}} \right) + Z[i]} \right|,}&{i = c,}\\
{\left| {Z[i]} \right|,}&{i \ne {I_2},c.}
\end{array}} \right.
\end{align}
Considering  properties of the complex Gaussian distribution, the distribution of $\left| {{{\ddot \Upsilon }_{i|c \ne {I_2}}^A}} \right|$ follows a Rayleigh distribution for ${i \ne {I_2},c}$ and a Rice distribution for ${i = {I_2}}$ and ${i = c}$, given by
\begin{align}
\label{eq:RIS_interferenceA_distrbution}
\left| {{{\ddot \Upsilon }_{i|c \ne {I_2}}^A}} \right| \sim \left\{ {\begin{array}{*{20}{l}}
{{f_{\mathrm{Ri} }}\left( {x;\left| {{H_A}} \right|\sqrt {{E_I}} {\chi _I},\frac{{{N_0}}}{2}} \right),}&{i = {I_2,}}\\
{{f_{\mathrm{Ri} }}\left( {x;\left| H \right|\sqrt {{E_c}}  ,\frac{{{N_0}}}{2}} \right),}&{i = c,}\\
{{f_{\mathrm{Ra} }}\left( {x;\frac{{{N_0}}}{2}} \right),}&{i \ne {I_2},c.}
\end{array}} \right.
\end{align}

Based on (\ref{eq:RIS_interferenceA_magnitude}), the symbol error rate (SER) of non-coherent detection in Case A is divided into noise-driven SER $P_{{A_{{\rm{NC}}}}}^{\rm{N}}$ and interference-driven SER $P_{{A_{{\rm{NC}}}}}^{\rm{I}}$, where  ``N" and ``I" mean noise-driven and interference-driven, respectively, and ``NC" represents non-coherent detection. The conditional noise-driven SER can be expressed as
\begin{align}
\label{eq:RIS_interferenceA_NCnoise_conditional}
P_{{A_{{\rm{NC}}}}}^{{\rm{N}},\left| H \right|} = {\rm{Pr}}\left( {\mathop {\max }\limits_{i \ne {I_2},c} \left| {Z[i]} \right| > \left| {\left| H \right|\sqrt {{E_c}} \exp \left( {j{\theta _{{h_{Td}}}}} \right) + Z[c]} \right|} \right).
\end{align}

According (\ref{eq:RIS_interferenceA_distrbution}), the event probability (\ref{eq:RIS_interferenceA_NCnoise_conditional}) requires to compute the maximum of $K-2$ noise-driven Rayleigh RVs large than a RV following Rice distribution of the transmitted symbol. An accurate closed-form approximation for the above probability event is provided in \cite{8392707}. Thus, the conditional noise-driven SER $P_{{A_{{\rm{NC}}}}}^\mathrm{N}$ can be obtained as
\begin{align}
\label{eq:RIS_interferenceA_NCnoise_condit_close}
P_{{A_{{\rm{NC}}}}}^\mathrm{N} \approx \int_0^\infty  {Q\left( {\sqrt {2x\Gamma K}  - \sqrt {2{\Phi _{K - 2}}} } \right){f_{{\left| H \right|^2}}}\left( x \right)dx} ,
\end{align}
where $\Gamma  = \frac{{{{{E_c}} \mathord{\left/{\vphantom {{{E_c}} {{T_c}}}} \right. \kern-\nulldelimiterspace} {{T_c}}}}}{{{N_0}B}} = \frac{{{E_c}}}{{{N_0}K}}$ is the average SNR, ${\Phi  _m} \approx \ln (m) + \frac{1}{{2m}} + 0.57722$ is the $m$-th harmonic number.

The next step is to obtain the distribution of the RV ${\left| {{H}} \right|^2}$ and solve the integral. We use the moment-matching method to fit the RV $\left| H \right|$ to the Gamma distribution, which is determined by the two parameters ${\omega _{H}}$ and ${\lambda _{H}}$, i.e., $\left| H \right|\sim{\rm{Gamma}}\left( {{\omega _H},{\lambda _H}} \right)$, where the estimators can be respectively expressed as
${\omega _H} = \left( {\frac{{{{\left[ {{\mu _{\left| H \right|}}\left( 1 \right)} \right]}^2}}}{{{\mu _{\left| H \right|}}\left( 2 \right) - {\mu _{\left| H \right|}}\left( 1 \right)}}} \right)$,
${\lambda _H} = \left( {\frac{{{\mu _{\left| H \right|}}\left( 1 \right)}}{{{\mu _{\left| H \right|}}\left( 2 \right) - {\mu _{\left| H \right|}}\left( 1 \right)}}} \right)$,
with ${{\mu _{\left| H \right|}}\left( 1 \right)}$ and ${{\mu _{\left| H \right|}}\left( 2 \right)}$ given in (\ref{H_1th}) and (\ref{H_2th}), respectively. The detailed derivations of ${\omega _H}$ and ${\lambda _H}$ are given in Appendix \ref{Hdistribution}. Then, the PDF of $\left| H \right|$ can be expressed as
\begin{align}
{f_{\left| H \right|}}\left( x \right) \approx \frac{{\lambda _H^{{\omega _H}}}}{{\gamma \left( {{\omega _H}} \right)}}{x^{{\omega _H} - 1}}\exp \left( { - {\lambda _H}x} \right),x \ge 0.
\end{align}
Furthermore, the PDF of ${\left| H \right|^2}$ can be obtained as
\begin{align}
\label{H2_PDF}
{f_{{{\left| H \right|}^2}}}\left( x \right) \approx \frac{1}{{2\sqrt x }}\frac{{\lambda _H^{{\omega _H}}}}{{\gamma \left( {{\omega _H}} \right)}}{\left( {\sqrt x } \right)^{{\omega _H} - 1}}\exp \left( { - {\lambda _H}\sqrt x } \right),x \ge 0.
\end{align}
$P_{{A_{{\rm{NC}}}}}^{\rm{N}}$ in (\ref{eq:RIS_interferenceA_NCnoise_condit_close}) is expressed as
\begin{align}
\label{eq:RIS_interferenceA_NCnoise_integral_conditional}
P_{{A_{{\rm{NC}}}}}^{\rm{N}} &\approx \int_0^\infty  Q \left( {\sqrt {2x\Gamma K}  - \sqrt {2{\Phi _{K - 2}}} } \right) \nonumber \\
&\times \frac{1}{{2\sqrt x }}\frac{{\lambda _H^{{\omega _H}}}}{{\gamma \left( {{\omega _H}} \right)}}{\left( {\sqrt x } \right)^{{\omega _H} - 1}}\exp \left( { - {\lambda _H}\sqrt x } \right)dx.
\end{align}
Let $t = \sqrt x$, and using the Gaussian $Q$-function approximation $Q(x) \approx \frac{1}{{12}}\exp \left( { - \frac{{{x^2}}}{2}} \right) + \frac{1}{4}\exp \left( { - \frac{{2{x^2}}}{3}} \right)$ \cite{4783779}, (\ref{eq:RIS_interferenceA_NCnoise_integral_conditional}) can be computed as
\begin{align}
\label{eq:RIS_interferenceA_NCnoise_integral1_conditional}
\begin{array}{l}
\!\!P_{{A_{{\rm{NC}}}}}^{\rm{N}} \approx \frac{{\lambda _H^{{\omega _H}}}}{{\gamma \left( {{\omega _H}} \right)}}\\
\!\! \times \!\!\left[ {\left. \begin{array}{l}
\frac{{\exp \left( { - {{U_2^2} \mathord{\left/
 {\vphantom {{U_2^2} 2}} \right.
 \kern-\nulldelimiterspace} 2}} \right)}}{{12}}\int_0^\infty  {\exp \!\left( { \!- \frac{{U_1^2}}{2}{t^2} \!-\! \left( {{\lambda _H}\! -\! {U_1}{U_2}} \right)t} \right){t^{{\omega _H} - 1}}dt} \\
\!\!\!\! + \frac{{\exp \left( { - {{2U_2^2} \mathord{\left/
 {\vphantom {{2U_2^2} 3}} \right.
 \kern-\nulldelimiterspace} 3}} \right)}}{4}\int_0^\infty  {\exp \!\left( {\! - \frac{{2U_1^2}}{3}{t^2} \!-\! \left( {{\lambda _H}\! -\! \frac{4}{3}{U_1}{U_2}} \right)t} \right){t^{{\omega _H} - 1}}dt}
\end{array} \!\!\right]\!\!,} \right.
\end{array}
\end{align}
where ${U_1} = \sqrt {2\Gamma K}$ and ${U_2} = \sqrt {2{\Phi _{K - 2}}}$. Utilizing (3.462.1) in \cite{zwillinger2007table}, (\ref{eq:RIS_interferenceA_NCnoise_integral1_conditional}) can be calculated as
\begin{align}
\label{eq:RIS_interferenceA_NCnoise_condit_close1}
P_{{A_{{\rm{NC}}}}}^{\rm{N}} \approx \lambda _H^{{\omega _H}}\left( {\frac{1}{{12}}{\Xi _1} + \frac{1}{4}{\Xi _2}} \right),
\end{align}
where ${\Xi _1}$ and ${\Xi _2}$ are, respectively, obtained as
\begin{align}
\label{eq:RIS_interferenceA_NCnoise_integral1_conditional_Xi _1}
&{\Xi _1} =U_1^{ - \omega_{H} }\!\exp \left( { - \frac{1}{2}U_2^2 + \frac{{{{\left( {\lambda  - {U_1}{U_2}} \right)}^2}}}{{4U_1^2}}} \right)\!{D\!    _{ - \omega_{H} }}\!\!\left(\!\! {\frac{{\lambda  - {U_1}{U_2}}}{{{U_1}}}} \!\right)\!,
\end{align} and
\begin{align}
{\Xi _2} \!\!=\!\! {\left(\!\! {\sqrt {\frac{4}{3}{U_1}} } \right)^{ \!\!\!\!- \omega_{H} }}\!\!\!\!\!\exp\!\! \left( {\!\! - \frac{2}{3}U_2^2 \!+\!\! \frac{{3{{\left( {\lambda  \!- \!\frac{4}{3}{U_1}{U_2}} \!\right)}^2}}}{{16U_1^2}}} \!\right)\!\!{D\!_{ \!- \omega_{H} }}\!\!\!\left( \!\!{\frac{{\lambda \!\! -\!\! \frac{4}{3}{U_1}{U_2}}}{{\sqrt {\frac{4}{3}{U_1}} }}} \!\!\right)\!\!.
\end{align}

Next, we derive the interference-driven probability $P_{{A_{{\rm{NC}}}}}^{\rm{I}}$. According to (\ref{eq:RIS_interferenceA_magnitude}), the conditional interference-driven probability $P_{{{A_{{\rm{NC}}}}}}^{\mathrm{I}|\begin{subarray}{l}
  \left| H \right|,\left| {{H_A}} \right| \\
   I ,\tau
\end{subarray} }$ can be expressed as
\begin{align}
\label{eq:RIS_interferenceA_NCinterfer_event}
\!P_{{\!{A_{{\rm{NC}}}}}}^{\mathrm{I}|\begin{subarray}{l}
  \left| H \right|,\left| {{H_A}} \right| \\
   I ,\tau
\end{subarray} }\!\!=
{\rm{Pr}}\left( \begin{array}{l}
\left| {\left| {{H_A}} \right|\sqrt {{E_I}} {\chi _I}\exp \left( {j\theta {_{H{_A}}}} \right) + Z[{I_2}]} \right|\\
 \ > \left| {\left| H \right|\sqrt {{E_c}} \exp \left( {j{\theta _{{h_{Td}}}}} \right) + Z[c]} \right|
\end{array} \right)\!.
\end{align}
Since (\ref{eq:RIS_interferenceA_distrbution}) shows that the magnitudes of the demodulation metric for $i = {I_2}$ and $i = c$ follow the Rice distribution, (\ref{eq:RIS_interferenceA_NCinterfer_event}) is attributed to the probabilistic comparison between two Rice RVs. The Rice distribution can be approximated as a Gaussian distribution for a larger Rice factor $\mathchar'26\mkern-10mu\lambda$. Thus, (\ref{eq:RIS_interferenceA_NCinterfer_event}) can be approximated as
\begin{align}
\label{eq:RIS_interferenceA_blind_appro}
P_{{{A_{{\rm{NC}}}}}}^{\mathrm{I}|\begin{subarray}{l}
 \left| H \right|,\left| {{H_A}} \right| \\
   I ,\tau
\end{subarray} } \approx Q\left( {\frac{{\sqrt {{{\left| {{H_A}} \right|}^2}{E_I}}   - \sqrt {{{\left| H \right|}^2}{E_c}} {\chi _I}}}{{\sqrt {{N_0}} }}} \right).
\end{align}

To obtain the interference-driven SER, we need to know the distribution of ${\left| {{H_A}} \right|^2}$. The distribution of ${\left| {{H_A}} \right|^2}$ can be fitted to the Gamma distribution using the central limit theorem (CLT) and moment-matching methods, i.e., ${\left| {{H_A}} \right|^2} \sim \left( {{\omega _A},{\lambda _A}} \right)$, where the estimators $\omega _A$ and ${\lambda _A}$ are given in (\ref{omigaA}) and (\ref{lamdaA}), respectively. The detailed derivations of ${\omega _A}$ and ${\lambda _A}$ are provided in Appendix \ref{Derivation_HA}. Hence, the PDF of ${{{\left| {{H_A}} \right|}^2}}$ is expressed as
\begin{align}
\label{pdf_HA2}
{f_{{{\left| {{H_A}} \right|}^2}}}\left( x \right) \approx \frac{{\lambda _A^{{\omega _A}}}}{{\gamma \left( {{\omega _A}} \right)}}{x^{{\omega _A} - 1}}\exp \left( { - {\lambda _A}x} \right),x \ge 0.
\end{align}

Using (\ref{H2_PDF}), (\ref{eq:RIS_interferenceA_blind_appro}), and (\ref{pdf_HA2}), one has
\begin{align}
\label{eq:err_inter_caseA_Non_coherent1}
P_{{A_{{\rm{NC}}}}}^{{\rm{I}}|I,\tau } &\approx \frac{{\lambda _H^{{\omega _H}}}}{{2\gamma \left( {{\omega _H}} \right)}}\frac{{\lambda _A^{{\omega _A}}}}{{\gamma \left( {{\omega _A}} \right)}}\int_0^\infty  {\int_0^\infty  {Q\left( {\frac{{\sqrt {x{E_s}}  - \sqrt {y{E_I}} {\chi _I}}}{{\sqrt {{N_0}} }}} \right)} }\nonumber \\&\times{\left( {\sqrt x } \right)^{{\omega _H} - 2}}{y^{{\omega _H} - 1}}\exp \left( { - {\lambda _H}\sqrt x  - {\lambda _A}y} \right)dxdy.
\end{align}
Due to the complexity of the double integral, we solve the integral by resorting to the Gauss-Hermite quadrature method \cite[Table 25.10]{zwillinger2007table}. $P_{{A_{{\rm{NC}}}}}^{{\rm{I}}|I,\tau }$ is calculated as
\begin{align}
\label{Interfererence_closefrom_caseA_Ncoherent}
P_{{A_{{\rm{NC}}}}}^{{\rm{I}}|I,\tau } &\approx \frac{{\lambda _H^{{\omega _H}}}}{{\gamma \left( {{\omega _H}} \right)}}\frac{{\lambda _A^{{\omega _A}}}}{{\gamma \left( {{\omega _A}} \right)}}\sum\limits_{{\upsilon _2} = 1}^{{V_2}} {\sum\limits_{{\upsilon _1} = 1}^{{V_1}} {{\psi _\upsilon }_{_2}{\psi _\upsilon }_{_1}} } \times \nonumber \\& Q\left[ {\sqrt {\Gamma K} \exp \left( {{\alpha _\upsilon }_{_1}} \right) - \sqrt {\Gamma K\exp \left( {{\beta _\upsilon }_{_1}} \right)} {\chi _I}} \right] \nonumber\\&\times \exp \left[ \begin{array}{l}
\alpha _{{\upsilon _1}}^2 + {\omega _H}{\alpha _\upsilon }_{_2} - {\lambda _H}\exp \left( {{\alpha _\upsilon }_2} \right)\\
 + \alpha _{{\upsilon _2}}^2 + {\omega _A}{\alpha _\upsilon }_1 - {\lambda _A}\exp \left( {{\alpha _\upsilon }_1} \right)
\end{array} \right].
\end{align}
The detailed derivation of (\ref{Interfererence_closefrom_caseA_Ncoherent}) is given in Appendix \ref{closed-form_Gauss_app}.

Given that $\tau $ and $I$ are uniformly distributed between $0$ and $\frac{K}{2}$ as well as $0$ and $K - 1$, respectively, the interference-driven probability can be expressed as
\begin{align}
\label{eq:caseA_inter_closeform3}
P_{{A_{{\rm{NC}}}}}^{\rm{I}} \approx \frac{1}{{K\left( {{K \mathord{\left/
 {\vphantom {K {2 + 1}}} \right.
 \kern-\nulldelimiterspace} {2 + 1}}} \right)}}\sum\limits_{\tau  = 0}^{{K \mathord{\left/
 {\vphantom {K 2}} \right.
 \kern-\nulldelimiterspace} 2}} {\sum\limits_{I = 0}^{K - 1} {P_{{A_{{\rm{NC}}}}}^{{\rm{I}}|I,\tau }} }.
\end{align}

Hence, combining (\ref{eq:RIS_interferenceA_NCnoise_condit_close1}) and (\ref{eq:caseA_inter_closeform3}), the BER for Case A with non-coherent detection is obtained as
\begin{align}
\label{eq:caseA_BER_closeform}
P_{{A_{{\rm{NC}}}}}^{BER} \approx \frac{{{K \mathord{\left/
 {\vphantom {K 2}} \right.
 \kern-\nulldelimiterspace} 2}}}{{K - 1}} \times \left[ {1 - \left( {1 - P_{{A_{{\rm{NC}}}}}^{\rm{I}}} \right)\left( {1 - P_{{A_{{\rm{NC}}}}}^{\rm{N}}} \right)} \right].
\end{align}

From (\ref{eq:caseA_BER_closeform}), one has $\frac{{{K \mathord{\left/ {\vphantom {K 2}} \right.
 \kern-\nulldelimiterspace} 2}}}{{K - 1}} \approx 0.5$, which reflects the fact that for each symbol SER, only half
of the underlying symbol bits are expected to be in error \cite{8392707}.
\subsubsection{Coherent detection} \label{sect:BER coherent detection_caseA}The BER for Case A with coherent detection will be analyzed in the following. Since the perfect CSI is adopted in this paper, it is feasible to get the compensated phase shift ${\theta _A} =  - {\theta _{Td}}$. According to (\ref{eq:RIS_free_inter_co_budengyu}) and (\ref{eq:int_yA1}), the coherent demodulation metric can be expressed as
\begin{align}
\label{eq:RIS_interferenceA_conherent_real_metric}
\Re \!\left( \!{{{\ddot \Upsilon }_{i|c \ne {I_2}}^A}\exp \left( {j{\theta _A}} \right)} \!\right) \!\approx \!\left\{ \!{\begin{array}{*{20}{l}}
{\!\left| {{H_A}} \right|\sqrt {{E_I}}  {\chi _I}\cos {\vartheta _A} + \eta [i],}\!&{i = {I_2,}}\\
{\!\left| H \right|\sqrt {{E_c}}   + \eta [i],}\!&{i = c,}\\
{\eta [i],}\!&{i \ne {I_2},c,}
\end{array}} \right.\,
\end{align}
where ${\vartheta _A} = {\theta _A} + {\theta _{{H_{Id}}}}$ is generally uniformly distributed in $[0,2\pi )$. The distribution of $\Re \left( {{{\ddot \Upsilon }_{i|c \ne {I_2}}^A}\exp \left( {j{\theta _A}} \right)} \right)$ can be obtained as
\begin{align}
\label{eq:RIS_interferenceA_conherent_real_metric_distribution}
\Re \!\left( \!{{{\ddot \Upsilon }_{i|c \ne {I_2}}^A}\exp \left( {j{\theta _A}} \right)} \right) \!\!\sim \!\!\left\{ {\begin{array}{*{20}{l}}
{\!\!{\cal N}\!\left( {\left| {{H_A}} \right|\sqrt {{E_I}}  {\chi _I}\cos {\vartheta \!_A},\!\frac{{{N_0}}}{2}} \right)\!\!,}&{\!\!i = {I_2,}}\\
{\!\!{\cal N}\!\left( {\left| H \right|\sqrt {{E_c}}  ,\frac{{{N_0}}}{2}} \right),}&{\!\!i = c,}\\
{\!\!{\cal N}\left( {0,\frac{{{N_0}}}{2}} \right),}&{\!\!i \ne {I_2},c.}
\end{array}} \right.
\end{align}
Similarly, the SER for Case A with coherent detection is divided into noise-driven SER $P_{{A_{{\rm{C}}}}}^{\rm{N}}$ and $P_{{A_{{\rm{C}}}}}^{\rm{I}}$, where ``C" mean coherent detection. From (\ref{eq:RIS_interferenceA_conherent_real_metric}) and (\ref{eq:RIS_interferenceA_conherent_real_metric_distribution}), the conditional noise-driven SER $P_{{A_{{\rm{C}}}}}^{{\rm{N}},\left| H \right|}$ can be expressed as
\begin{align}
\label{eq:RIS_interferenceA_conherent_noise_event}
P_{{A_{{\rm{C}}}}}^{{\rm{N}},\left| H \right|} = {\rm{Pr}}\left( {\mathop {\max }\limits_{i \ne {I_2},c} \eta [i] > \sqrt {{{\left| H \right|}^2}{E_c}}  + \eta [c]} \right),
\end{align}
where the probability event occurs when the maximum of $K-2$ noise-driven Gaussian random variables is larger than the Gaussian random variable of the transmitted symbol. \cite{8835951} provides a low-complexity approximation to computed the above probability event for $SF \ge 7$. Thus, the noise-driven SER for Case A with coherent detection can be expressed as
\begin{align}
\label{eq:RIS_interferenceA_conherent_integral}
P_{{A_{{\rm{C}}}}}^{\rm{N}} &\approx \int_0^\infty   Q\left( {\frac{{\sqrt {\Gamma Kx}  - {\varepsilon _1}}}{{{\varepsilon _2}}}} \right)\nonumber\\& \times \frac{1}{2}\frac{{\lambda _H^{{\omega _H}}}}{{\gamma \left( {{\omega _H}} \right)}}{\left( {\sqrt x } \right)^{{\omega _H} - 2}}\exp \left( { - {\lambda _H}\sqrt x } \right)dx,
\end{align}
where ${\varepsilon _1} = \sqrt {{\textstyle{1 \over 2}}} \left( {1.161 + 0.2074SF} \right)$, and ${\varepsilon _2} = \sqrt {{\textstyle{1 \over 2}} + {\textstyle{1 \over 2}}\left( {0.2775 - 0.0153SF} \right)} $. Following the similar approach as in (\ref{eq:RIS_interferenceA_NCnoise_condit_close1}), (\ref{eq:RIS_interferenceA_conherent_integral}) can be computed as
\begin{align}
\label{eq:RIS_interferenceA_conherent_integral_closeform}
P_{{A_{{\rm{C}}}}}^{\rm{N}} \approx \frac{1}{{12}}{\Xi _3} + \frac{1}{4}{\Xi _4},
\end{align}
where
\begin{align}
&{\!\Xi _3}\!\! =\!\! {\left(\!\! {\frac{{{\lambda _H}}}{{\sqrt {\Gamma K} }}} \right)^{\!\!{\omega _H}}}\!\!\!\!\!\exp \!\!\left(\!\!\! { - \frac{{\varepsilon _2^2}}{{2\varepsilon _1^2}}\! \!+\!\! \frac{{{{\!\left( {{\varepsilon _1}{\lambda _H} \!\!-\!\! \frac{{\sqrt {\Gamma K} {\varepsilon _2}}}{{2{\varepsilon _1}}}} \!\!\right)}^2}}}{{4\sqrt {\Gamma K} }}} \!\right)\!\!{D\!_{ - {\omega \!_H}}}\!\!\left(\!\! {\frac{{{\varepsilon _1}{\lambda _H}}}{{\sqrt {\Gamma K} }} \!\!-\!\! \frac{{{\varepsilon _2}}}{{2{\varepsilon _1}}}} \!\right)\!\!,
\end{align}
and
\begin{align}
&\!{\Xi _4} \!\!=\!\!\! {\left(\!\! {\sqrt {\frac{\!\!\!{3\varepsilon _1^2}}{\!\!{4\Gamma K}}} } \right)^{\!\!{\omega _H}}}\!\!\!\!\!\!\exp \!\!\left( \!\!\!{ - \frac{{4\varepsilon _2^2}}{{3\varepsilon _1^2}}\! +\! \frac{\!{3\varepsilon _1^2{\lambda _H} \!-\! 4\sqrt {\Gamma K} {\varepsilon _2}}}{\!{16\Gamma K}}}\!\!\! \right)\!\!\!{D\!_{ - {\omega _H}}}\!\!\!\left(\! {\frac{\!{{\varepsilon _1}{\lambda _H}}}{\!{\sqrt {\Gamma K} }}\!\! -\!\! \frac{{{\varepsilon _2}}}{{2{\varepsilon _1}}}} \!\right)\!\!.
\end{align}

From (\ref{eq:RIS_interferenceA_conherent_real_metric}), the conditional interference-driven symbol error probability for coherent detection can be expressed as
\begin{align}
\label{eq:RIS_interferenceA_Cinterfer_condition_event}
\begin{array}{l}
\!\!\!\!P_{{{A_{{\rm{C}}}}}}^{\mathrm{I}|\begin{subarray}{l}
  \left| H \right|,\left| {{H_A}} \right|,{\vartheta_{A}}  \\
   I ,\tau
\end{subarray} }\!\!\!\!\!=
{\rm{Pr}}\!\left( {\left| {{H_A}} \right|\sqrt {{E_I}}  {\chi _I}\cos {\vartheta_{A}}\! +\! \eta[{I_2}] \!>\!  {\left| H \right|\sqrt {{E_c}} \! +\! \eta[c]} } \right)\!\!.
\end{array}
\end{align}
Furthermore, using (\ref{eq:RIS_interferenceA_conherent_real_metric_distribution}), the interference-driven probability $P_{{A_{\rm{C}}}}^{{\rm{I}}|I,\tau }$ can be obtained as
\begin{align}
\label{eq:RIS_interferenceA_Cinterfer_integral_event}
\!\!P_{{A_{\rm{C}}}}^{{\rm{I}}|I,\tau } \!&\!\approx\! \frac{1}{{4\pi }}\frac{{\lambda _H^{{\omega _H}}}}{{\gamma \left( {{\omega _H}} \right)}}\frac{{\lambda _A^{{\omega _A}}}}{{\gamma \left( {{\omega _A}} \!\right)}}\!\!\int_0^{2\pi }\! \!\!\int_0^\infty \! \! \!\!\int_0^\infty   \!\!Q\!\left(\! {\frac{{\sqrt {x{E_s}} \! -\! \sqrt {y{E_I}} \cos \vartheta {\chi _I}}}{{\sqrt {{N_0}} }}}\!\! \right) \nonumber\\
&\times {x^{\frac{{{\omega _H} - 2}}{2}}}{y^{{\omega _A} - 1}}\exp \left( { - {\lambda _H}\sqrt x  - {\lambda _H}_{_A}y} \right)dxdyd\vartheta.
\end{align}
Then, using the Gauss-Hermite quadrature and staircase approximation methods, (\ref{eq:RIS_interferenceA_Cinterfer_integral_event}) can be computed as
\begin{align}
\label{eq:RIS_interferenceA_Cinterfer_closeform_event}
P_{{A_{\rm{C}}}}^{{\rm{I}}|I,\tau } &\approx \frac{1}{M}\frac{{\lambda _H^{{\omega _H}}}}{{\gamma \left( {{\omega _H}} \right)}}\frac{{\lambda _A^{{\omega _A}}}}{{\gamma \left( {{\omega _A}} \right)}}\sum\limits_{\varsigma = 1}^M {\sum\limits_{{\upsilon _2} = 1}^{{V_2}} {\sum\limits_{{\upsilon _1} = 1}^{{V_1}} {{\psi _\upsilon }_{_2}{\psi _\upsilon }_{_1}} }}\nonumber\\ &\times
 Q\left[ {\sqrt {\Gamma K} \exp \left( {{\alpha _\upsilon }_{_1}} \right) - \sqrt {\Gamma K\exp \left( {{\beta _\upsilon }_{_2}} \right)} \cos \frac{{2\pi \varsigma}}{M}{\chi _I}} \right]\nonumber\\ &\times\exp \left[ \begin{array}{l}
\alpha _{{\upsilon _1}}^2 + {\omega _H}{\alpha _\upsilon }_{_1} - {\lambda _H}\exp \left( {{\alpha _\upsilon }_{_1}} \right)\\
 + \beta _{{\upsilon _2}}^2 + {\omega _A}{\beta _\upsilon }_{_2} - {\lambda _A}\exp \left( {{\beta _\upsilon }_{_2}} \right)
\end{array} \right] .
\end{align}
The detailed proof is provided in Appendix \ref{closed-form_Gauss_app}. Hence, the interference-driven SER $P_{{A_{{\rm{NC}}}}}^{\rm{I}}$
can be expressed as
\begin{align}
\label{caseA_Coherent_interdriven_closeform1}
P_{{A_{{\rm{C}}}}}^{\rm{I}} \approx \frac{1}{{K\left( {{K \mathord{\left/
 {\vphantom {K 2}} \right.
 \kern-\nulldelimiterspace} 2} + 1} \right)}}\mathop \sum \limits_{\tau  = 0}^{{K \mathord{\left/
 {\vphantom {K 2}} \right.
 \kern-\nulldelimiterspace} 2}} \mathop \sum \limits_{I = 0}^{K - 1} P_{{A_{{\rm{C}}}}}^{{\rm{I}}|I,\tau }.
\end{align}
Combining (\ref{eq:RIS_interferenceA_conherent_integral_closeform}) and (\ref{caseA_Coherent_interdriven_closeform1}), the BER for Case A with coherent detection can be obtained as
\begin{align}
\label{caseA_coherent_total_BER}
P_{{A_{\rm{C}}}}^{BER} \approx \frac{{{K \mathord{\left/
 {\vphantom {K 2}} \right.
 \kern-\nulldelimiterspace} 2}}}{{K - 1}} \times \left[ {1 - \left( {1 - P_{{A_{{\rm{C}}}}}^{\rm{I}}} \right)\left( {1 - P_{{A_{{\rm{C}}}}}^{\rm{N}}} \right)} \right].
\end{align}

{{\em Remark~2:}
To provide further insights, we use the second moments ${\mu _{{J_T}}}\left( 2 \right) = N  {\mu _{{\zeta _{T\ell }}}}\left( 2 \right) + N  \left( {N - 1} \right) {\left[ {{\mu _{{\zeta _{T\ell }}}}\left( 1 \right)} \right]^2}$ (See Appendix \ref{Hdistribution}) and
${\mu _{\left| G \right|}}\left( 2  \right) = N$ (See Appendix \ref{Derivation_HA}) to evaluate the gains from the RIS in ${\left| {{H}} \right|^2}$ and ${\left| {{H_A}} \right|^2}$, respectively. These show that the average received power of the target user and the interfering user are quadratically and linearly enhanced as $N$ increases, respectively. Hence, increasing the number of reflecting elements can suppress the interfering signal,
which is consistent with the result in Fig.~\ref{fig:diffREBER} (a).
\subsection{BER Analysis for Case B} \label{sect:BER CaseB}
\subsubsection{Non-coherent detection} \label{sect:BER non_coherentCaseB}
From (\ref{eq:int_yB2}), the non-coherent demodulation metric for Case B can be expressed as
\begin{align}
\label{eq:demo_caseB_metric}
\left| {{\ddot \Upsilon _{i|c \ne {I_2}}^B}} \right|  \approx \left\{ {\begin{array}{*{20}{l}}
{\left| {\left| {{H_B}} \right|\sqrt {{E_I}} {\chi _I}\exp \left( {j{\theta _{{H_{Id}}}}} \right) + Z[i]} \right|,}&{i = {I_2},}\\
{\left| {\left| H \right|\sqrt {{E_c}} \exp \left( {j{\theta _{{h_{Td}}}}} \right) + Z[i]} \right|,}&{i = c,}\\
{\left| {Z[i]} \right|,}&{i \ne {I_2},c.}
\end{array}} \right.
\end{align}
The distribution of $\left| {\ddot \Upsilon _{i|c \ne {I_2}}^B} \right|$ can be expressed as
\begin{align}
\label{CaseB_NC_metric_distri}
\left| {\ddot \Upsilon _{i|c \ne {I_2}}^B} \right| \sim\left\{ {\begin{array}{*{20}{l}}
{{f_{{\rm{Ri}}}}\left( {x;\left| {{H_B}} \right|\sqrt {{E_I}} {\chi _I},\frac{{{N_0}}}{2}} \right),}&{i = {I_2,}}\\
{{f_{{\rm{Ri}}}}\left( {x;\left| H \right|\sqrt {{E_c}} ,\frac{{{N_0}}}{2}} \right),}&{i = c,}\\
{{f_{{\rm{Ra}}}}\left( {x;\frac{{{N_0}}}{2}} \right),}&{i \ne {I_2},c.}
\end{array}} \right.
\end{align}
Thus, the SER for Case B is also divided into noise-driven SER $P_{{B_{{\rm{NC}}}}}^{\rm{N}}$ and interference-driven SER $P_{{B_{{\rm{NC}}}}}^{\rm{I}}$.
Similar to the derivation of (\ref{eq:RIS_interferenceA_NCnoise_conditional}) and (\ref{eq:RIS_interferenceA_NCnoise_condit_close1}), the conditional noise-driven SER for Case B can be obtained as
\begin{align}
\label{eq:RIS_interferenceB_NCnoise_condit_close1}
P_{{B_{{\rm{NC}}}}}^{\rm{N}} \approx \lambda _H^{{\omega _H}}\left( {\frac{1}{{12}}{\Xi _1} + \frac{1}{4}{\Xi _2}} \right).
\end{align}

Since in Case B, the RIS controllers can obtain the target and interfering users' CSIs, the distribution of $\left| {{H_B}} \right|$ can be approximated to the Gamma distribution by using the same moment estimation step of $\left| {{H}} \right|$. The PDF of $\left| {{H_B}} \right|$ is given by
\begin{align}
\label{HB_PDF}
{f_{\left| {{H_B}} \right|}}\left( x \right) \approx \frac{{\lambda _H^{{\omega _H}}}}{{\gamma \left( {{\omega _H}} \right)}}{x^{{\omega _H} - 1}}\exp \left( { - {\lambda _H}x} \right),x \ge 0.
\end{align}

According to (\ref{CaseB_NC_metric_distri}), for large Rice factors, the conditional interference-driven SER for Case B can be computed as
\begin{align}
\label{CaseB_NC_integral}
P_{{B_{{\rm{NC}}}}}^{{\rm{I}}|I,\tau } &\approx \frac{{\lambda _H^{2{\omega _H}}}}{{4{{\left[ {\gamma \left( {{\omega _H}} \right)} \right]}^2}}}\int_0^\infty  {\int_0^\infty  {Q\left( {\frac{{\sqrt {x{E_c}}  - \sqrt {y{E_I}} {\chi _I}}}{{\sqrt {{N_0}} }}} \right)} } \nonumber\\ & \times {\left( {\sqrt x } \right)^{{\omega _H} - 2}}{\left( {\sqrt y } \right)^{{\omega _H} - 2}}\exp \left( { - {\lambda _H}\sqrt x  - {\lambda _H}\sqrt y } \right).
\end{align}
Using the Gauss-Hermite quadrature method, (\ref{CaseB_NC_integral}) can be calculated as
\begin{align}
P_{{B_{{\rm{NC}}}}}^{{\rm{I}}|I,\tau } &\approx \frac{{\lambda _H^{2{\omega _H}}}}{{{{\left[ {\gamma \left( {{\omega _H}} \right)} \right]}^2}}}\sum\limits_{{\upsilon _2} = 1}^{{V_2}} {\sum\limits_{{\upsilon _1} = 1}^{{V_1}} {{\upsilon _2}{\upsilon _1} }}\nonumber\\
& \times Q\left( {\sqrt {\Gamma K} \exp \left( {{x_{\upsilon 2}}} \right) - \sqrt {\Gamma K} \exp \left( {{x_{\upsilon 1}}} \right){\chi _I}} \right)\nonumber\\
& \times {\exp \left\{ \begin{array}{l}
x_{{\upsilon _2}}^2 +  + x_{{\upsilon _1}}^2 + {\omega _H}\left( {{x_{{\upsilon _2}}} + {x_{{\upsilon _1}}}} \right)\\
 - {\lambda _H}\left[ {\exp \left( {{x_{{\upsilon _2}}}} \right) + \exp \left( {{x_{{\upsilon _1}}}} \right)} \right]
\end{array} \right\}}.
\end{align}
The detailed proof is provided in Appendix \ref{closed-form_Gauss_app}.

Thus, $P_{{B_{{\rm{NC}}}}}^{\rm{I}}$ can be obtained as
\begin{align}
\label{eq:closeform_caseB_NC}
P_{{B_{{\rm{NC}}}}}^{\rm{I}} \approx \frac{1}{{K\left( {{K \mathord{\left/
 {\vphantom {K 2}} \right.
 \kern-\nulldelimiterspace} 2} + 1} \right)}}\mathop \sum \limits_{\tau  = 0}^{{K \mathord{\left/
 {\vphantom {K 2}} \right.
 \kern-\nulldelimiterspace} 2}} \mathop \sum \limits_{I = 0}^{K - 1} P_{{B_{{\rm{NC}}}}}^{{\rm{I}}|I,\tau }.
\end{align}
Combining (\ref{eq:RIS_interferenceB_NCnoise_condit_close1}) and (\ref{eq:closeform_caseB_NC}), the BER for Case B with non-coherent detection can be obtained as
\begin{align}
\label{caseB_NC_totalBER}
P_{{B_{{\rm{NC}}}}}^{BER} \approx \frac{{{K \mathord{\left/
 {\vphantom {K 2}} \right.
 \kern-\nulldelimiterspace} 2}}}{{K - 1}} \times \left[ {1 - \left( {1 - P_{{B_{{\rm{NC}}}}}^{\rm{I}}} \right)\left( {1 - P_{{B_{{\rm{NC}}}}}^{\rm{N}}} \right)} \right].
\end{align}
\subsubsection{Coherent detection} \label{sect:BER coherent detection_caseA}
Adhering to the previous analysis of coherent detection, we firstly rotate the phase of the target signal to obtain the coherent detection metric, given by
\begin{align}
\Re \!\left( {\ddot \Upsilon _{i|c \ne {I_2}}^B\exp \left( {j{\theta _B}} \right)} \right)\! \approx\! \left\{\!\! {\begin{array}{*{20}{l}}
{\left| {{H_B}} \right|\sqrt {{E_I}} {\chi _I}\cos {\vartheta _B} \!+ \!\eta [i],}&{\!i = {I_2,}}\\
{\left| H \right|\sqrt {{E_c}}  + \eta [i],}&{\!i = c,}\\
{\ \eta [i],}&{\!i \ne {I_2},c,}
\end{array}} \right.{\mkern 1mu}
\end{align}
where ${\theta _B} =  - {\theta _{{h_{Td}}}}$, and ${\vartheta _B} = {\mkern 1mu} {\theta _B} + {\mkern 1mu} {\theta _{{h_{Id}}}}$. The distribution of $\Re \left( {\ddot \Upsilon _{i|c \ne {I_2}}^B\exp \left( {j{\theta _B}} \right)} \right)$ can be obtained as
\begin{align}
\label{caseB_coherent_demo_metric_detri}
\Re\! \left(\! {\ddot \Upsilon _{i|c \ne {I_2}}^B\!\exp \left( {j{\theta _B}} \right)} \!\right)\!\sim\!\!\left\{ {\begin{array}{*{20}{l}}
{\!\!{\cal N}\!\left( {\left| {{H_B}} \right|\sqrt {{E_I}} {\chi _I}\cos {\vartheta _B},\frac{{{N_0}}}{2}} \right)\!,}&\!{\!i \!= \!{I_2},}\\
{\!\!{\cal N}\!\left( {\left| H \right|\sqrt {{E_c}} ,\frac{{{N_0}}}{2}} \right)\!,}&\!{\!i\! =\! c,}\\
{\!\!{\cal N}\left( {0,\frac{{{N_0}}}{2}} \right),}&{\!\!i \!\ne\! {I_2},c.}
\end{array}} \right.
\end{align}

According to (\ref{caseB_coherent_demo_metric_detri}), the SER for Case B with coherent detection can be divided into noise-driven SER $P_{{B_{{\rm{C}}}}}^{\rm{I}}$ and interference-driven SER $P_{{B_{{\rm{C}}}}}^{\rm{N}}$.
Following the same approach for (\ref{eq:RIS_interferenceA_conherent_noise_event}) and (\ref{eq:RIS_interferenceA_conherent_integral_closeform}), the noise-driven SER $P_{{B_{{\rm{C}}}}}^{\rm{N}}$ can be obtained as
\begin{align}
\label{caseB_noise_closeform_Coherent}
P_{{B_{{\rm{C}}}}}^{\rm{N}} \approx \frac{1}{{12}}{\Xi _3} + \frac{1}{4}{\Xi _4}.
\end{align}

Since $\left| H \right|$ and $\left| {{H_B}} \right|$ are approximated to the Gamma distribution, and ${\vartheta _B}$ is uniformly distributed over $\left[ {0,2\pi } \right)$, the conditional interference-driven SER for Case B with coherent detection can be expressed as
\begin{align}
\label{caseB_coherent_integral_inter}
P_{{B_{{\rm{C}}}}}^{{\rm{I}}|I,\tau }& \approx \frac{1}{{8\pi }}\frac{{\lambda _H^{2{\omega _H}}}}{{{{\left[ {\gamma \left( {{\omega _H}} \right)} \right]}^2}}}\!\!\int_0^{2\pi }\!\!\!\! {\int_0^\infty \!\!\!\! {\int_0^\infty  \!\!{Q\left( {\frac{{\sqrt {x{E_s}}  - \sqrt {y{E_I}} \cos \vartheta {\chi _I}}}{{\sqrt {{N_0}} }}} \!\right)} }}\nonumber
\\&\times{{{\left( {\sqrt {xy} } \right)}^{{\omega _H} - 2}}\exp \left( { - {\lambda _H}\sqrt x  - {\lambda _H}\sqrt y } \right)dxdy}\vartheta.
\end{align}
The closed-form expression of $P_{{B_{{\rm{C}}}}}^{{\rm{I}}|I,\tau }$ can be calculated as
\begin{align}
\label{caseB_coherent_inter_condition_closeform}
P_{{B_{\rm{C}}}}^{{\rm{I}}|I,\tau } &\approx \frac{1}{M}\frac{{\lambda _H^{2{\omega _H}}}}{{{{\left[ {\gamma \left( {{\omega _H}} \right)} \right]}^2}}}\sum\limits_{\varsigma  = 1}^M {\sum\limits_{{\upsilon _2} = 1}^{{V_2}} {\sum\limits_{{\upsilon _1} = 1}^{{V_1}} {{\psi _\upsilon }_{_2}{\psi _\upsilon }_{_1}} }}\nonumber\\
&\times Q\left[ {\sqrt {\Gamma K} \exp \left( {{\alpha _\upsilon }_{_1}} \right) - \sqrt {\Gamma K} \exp \left( {{\beta _\upsilon }_{_2}} \right)\cos \frac{{2\pi \varsigma }}{M}{\chi _I}} \right]\nonumber\\
& \times \exp \left\{ \begin{array}{l}
\alpha _{{\upsilon _1}}^2 + \beta _{{\upsilon _2}}^2 + {\omega _H}\left( {{\alpha _\upsilon }_1 + {\beta _\upsilon }_{_2}} \right)\\
 - {\lambda _H}\left[ {\exp \left( {{\alpha _\upsilon }_1} \right) + \exp \left( {{\beta _\upsilon }_2} \right)} \right]
\end{array} \right\}.
\end{align}
The detailed proof is provided in Appendix \ref{closed-form_Gauss_app}. Thus, the interference-driven SER for Case B with coherent detection can be obtained as
\begin{align}
\label{caseB_coherent_inter_closeform}
P_{{B_{\rm{C}}}}^{\rm{I}} \approx \frac{1}{{K\left( {{K \mathord{\left/
 {\vphantom {K 2}} \right.
 \kern-\nulldelimiterspace} 2} + 1} \right)}}\mathop \sum \limits_{\tau  = 0}^{{K \mathord{\left/
 {\vphantom {K 2}} \right.
 \kern-\nulldelimiterspace} 2}} \mathop \sum \limits_{I = 0}^{K - 1} P_{{B_{\rm{C}}}}^{{\rm{I}}|I,\tau }.
\end{align}
Finally, combining (\ref{caseB_noise_closeform_Coherent}) and (\ref{caseB_coherent_inter_closeform}), the BER for Case B with coherent detection is given by
\begin{align}
\label{caseB_coherent_BER_closeform}
P_{{B_{\rm{C}}}}^{BER} \approx \frac{{{K \mathord{\left/
 {\vphantom {K 2}} \right.
 \kern-\nulldelimiterspace} 2}}}{{K - 1}} \times \left[ {1 - \left( {1 - P_{{B_{\rm{C}}}}^{\rm{I}}} \right)\left( {1 - P_{{B_{\rm{C}}}}^{\rm{N}}} \right)} \right].
\end{align}
\section{Results and Discussions} \label{sect:Results and Discussions}
In this section, we show the BER performance of the proposed system (i.e., Case A and Case B) over Nakagami-$m$ fading channels. Since the RIS is usually deployed on high buildings, we assume that all communication links have a LOS component. To this end, we set the shape parameters as ${m_1} = {m_2} = {m_{{h_T}}} = {m_{{g_T}}} = {m_{{h_I}}} = {m_{{g_I}}} = m$ ($m>1$).\footnote{ The correlation between the shape parameter $m$ and Rician factor $\mathchar'26\mkern-10mu\lambda$ can be obtained as
$m = {\left( {1 - {{\left( {\frac{\mathchar'26\mkern-10mu\lambda }{{1 + \mathchar'26\mkern-10mu\lambda }}} \right)}^2}} \right)^{ - 1}}$ \cite{9406837}. When $m > 1$, we assume that the link has a LOS propagation condition ($m=1$ is the Rayleigh channel that corresponds to the NLOS environment). Moreover, larger values of $m$ will result in a stronger LOS condition.} The bandwidth $B$ is set to $125$ kHz.

\begin{figure}[htbp]
 \begin{center}
  \subfigure[]
  {
   \centering
   \includegraphics[width=3.8in,height=2.4in]{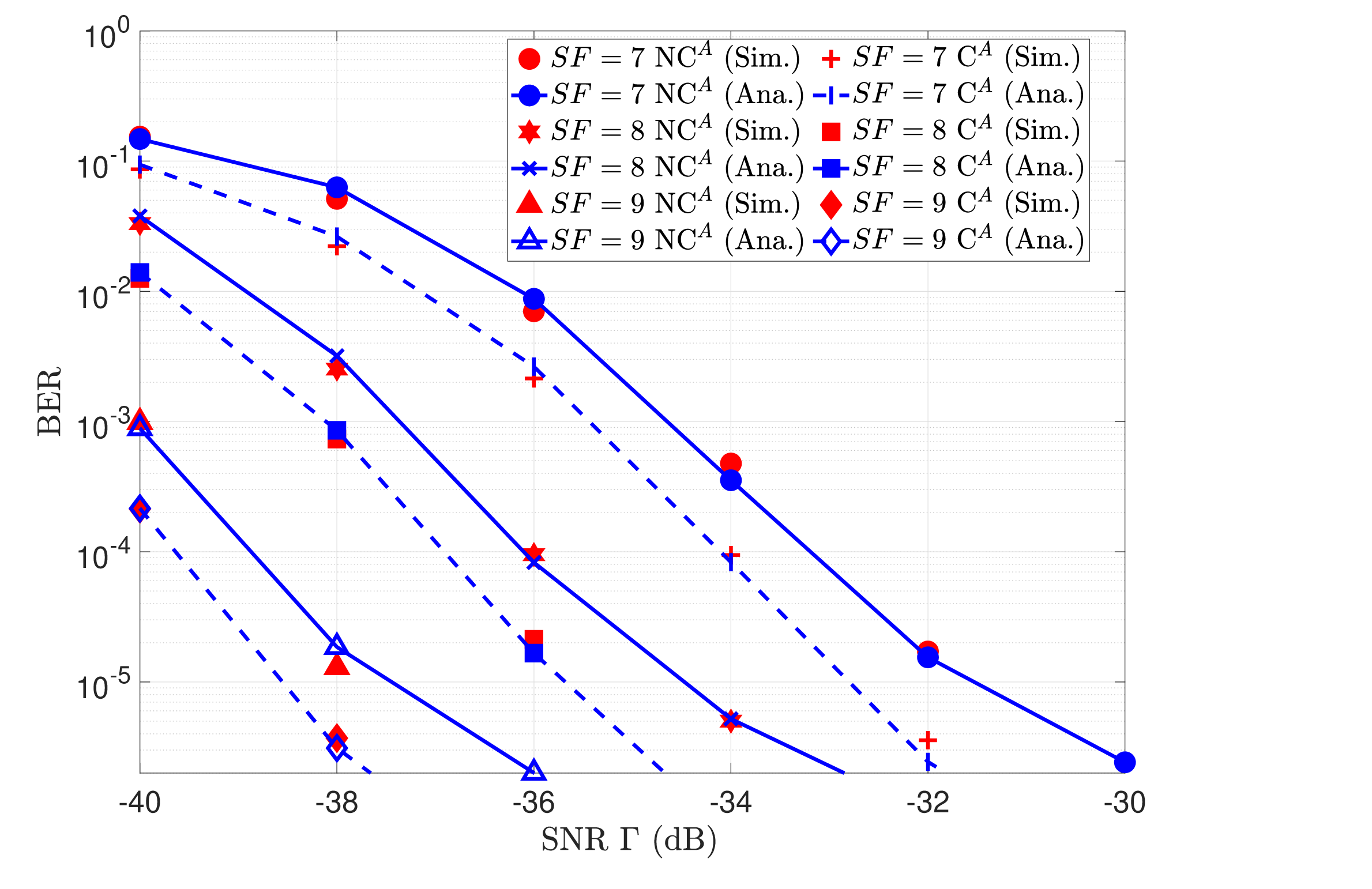}
   \label{BER_case_A}
  }
  \subfigure[]
  {
   \centering
   \includegraphics[width=3.8in,height=2.4in]{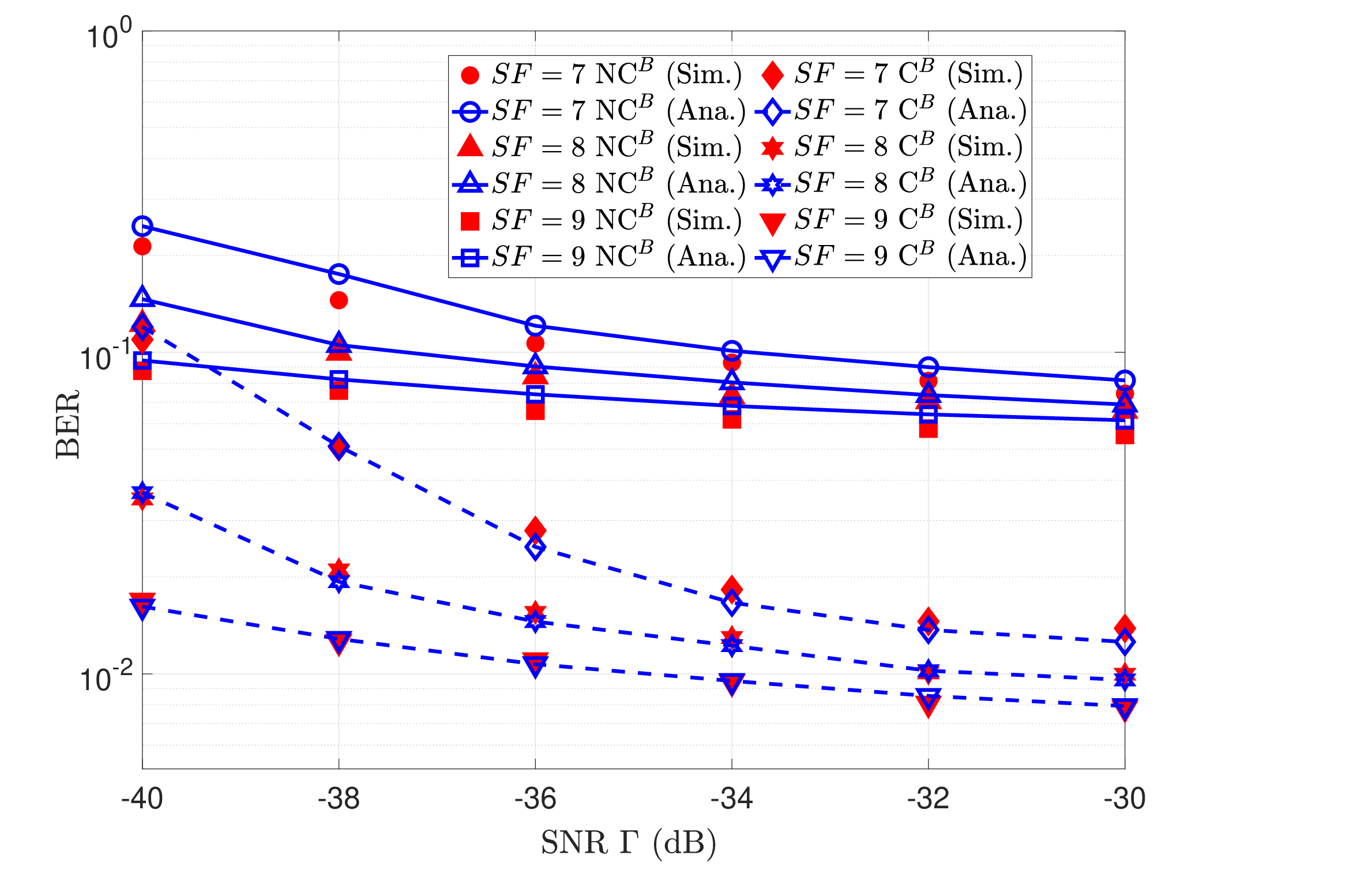}
   \label{BER_case_B}
  }
  \vspace{-0.2cm}
  \caption{Analytical and simulated BER for (a) Case A, and (b) Case B, where $SF=7,8,9$, $N=25$, and $m = 2$.} \label{fig:BER}
 \end{center}
\end{figure}
\begin{figure}[htbp]
 \begin{center}
  \subfigure[]
  {
   \centering
   \includegraphics[width=3.8in,height=2.5in]{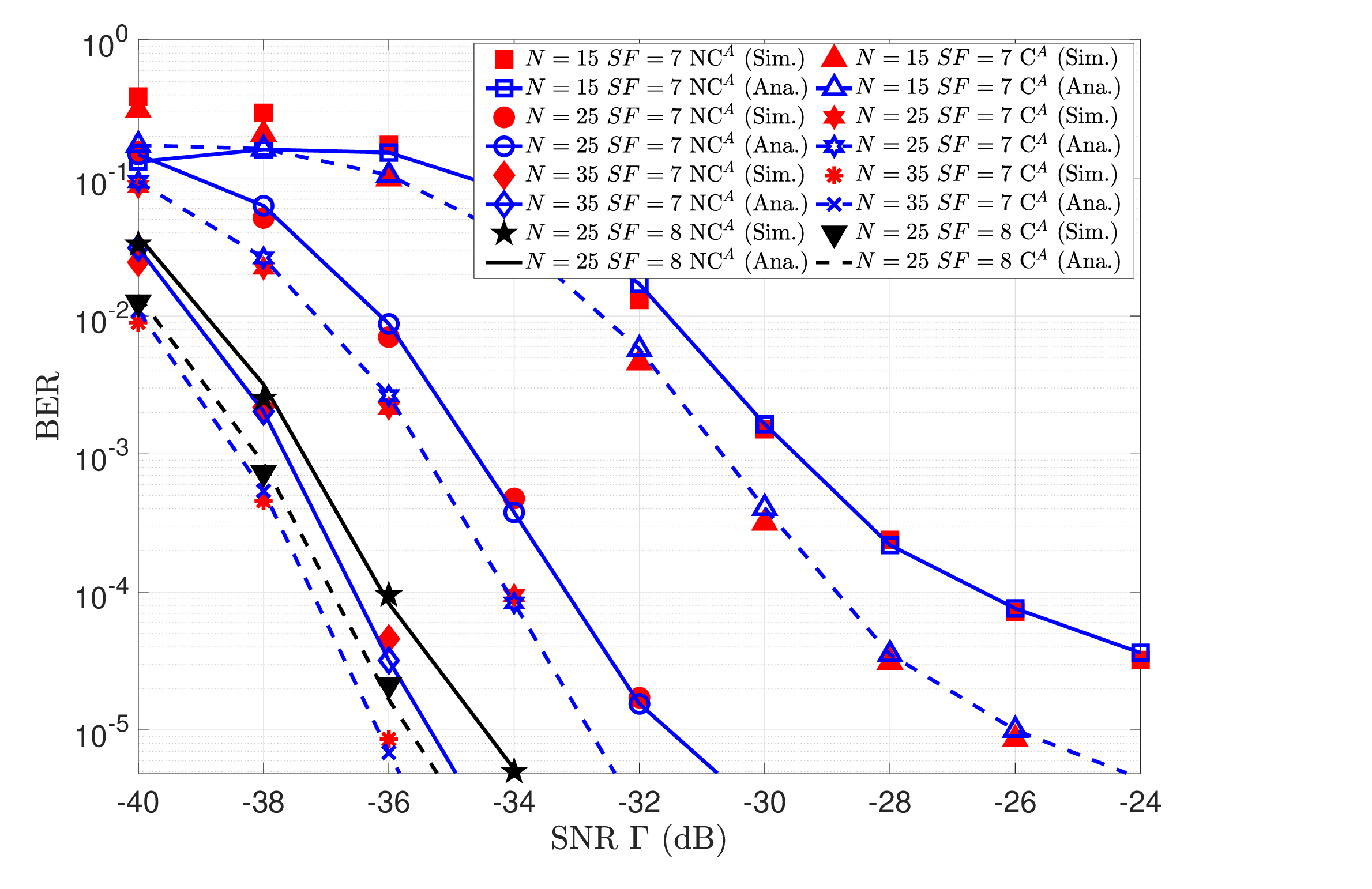}
   \label{BER_case_A}
  }
  \subfigure[]
  {
   \centering
   \includegraphics[width=3.8in,height=2.5in]{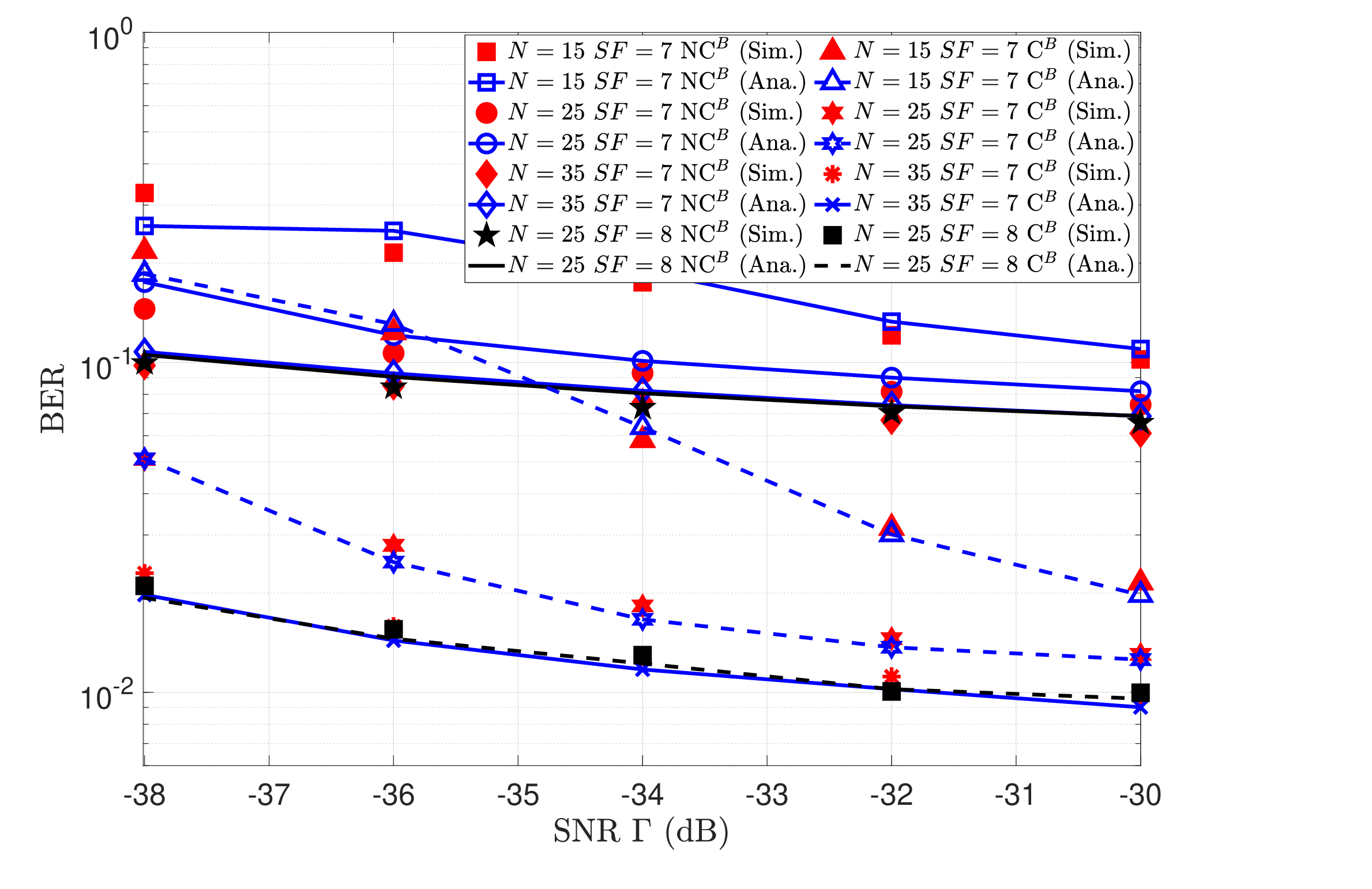}
   \label{BER_case_B}
  }
  \vspace{-0.2cm}
  \caption{Analytical and simulated BER curves for (a) Case A, and (b) Case B, with various values of $N$, where $SF=7$ and $m=2$.} \label{fig:diffREBER}
 \end{center}
\end{figure}
\begin{figure}[htbp]
 \begin{center}
  \subfigure[]
  {
   \centering
   \includegraphics[width=3.8in,height=2.5in]{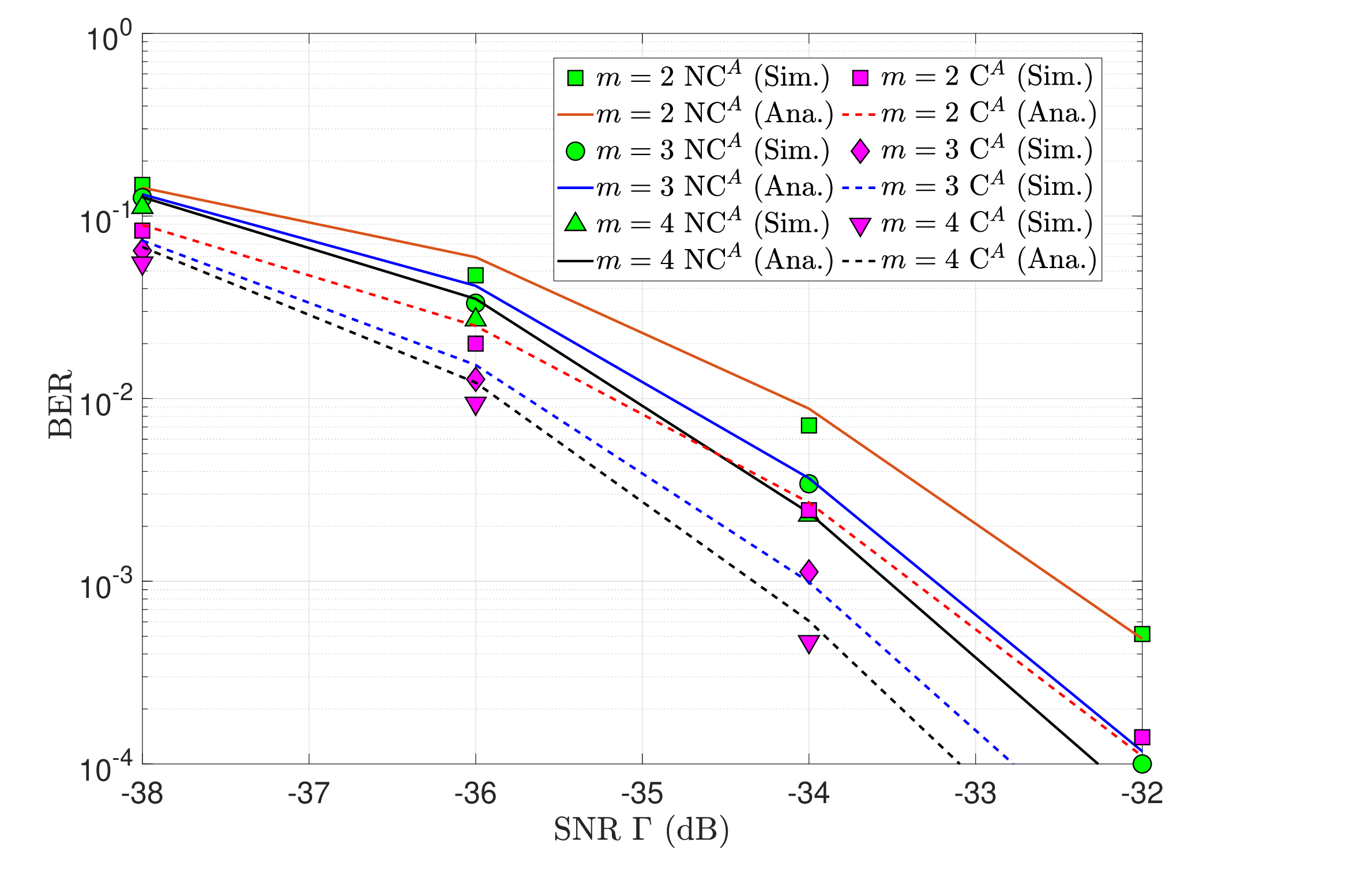}
   \label{BER_case_A}
  }
  \subfigure[]
  {
   \centering
   \includegraphics[width=3.8in,height=2.5in]{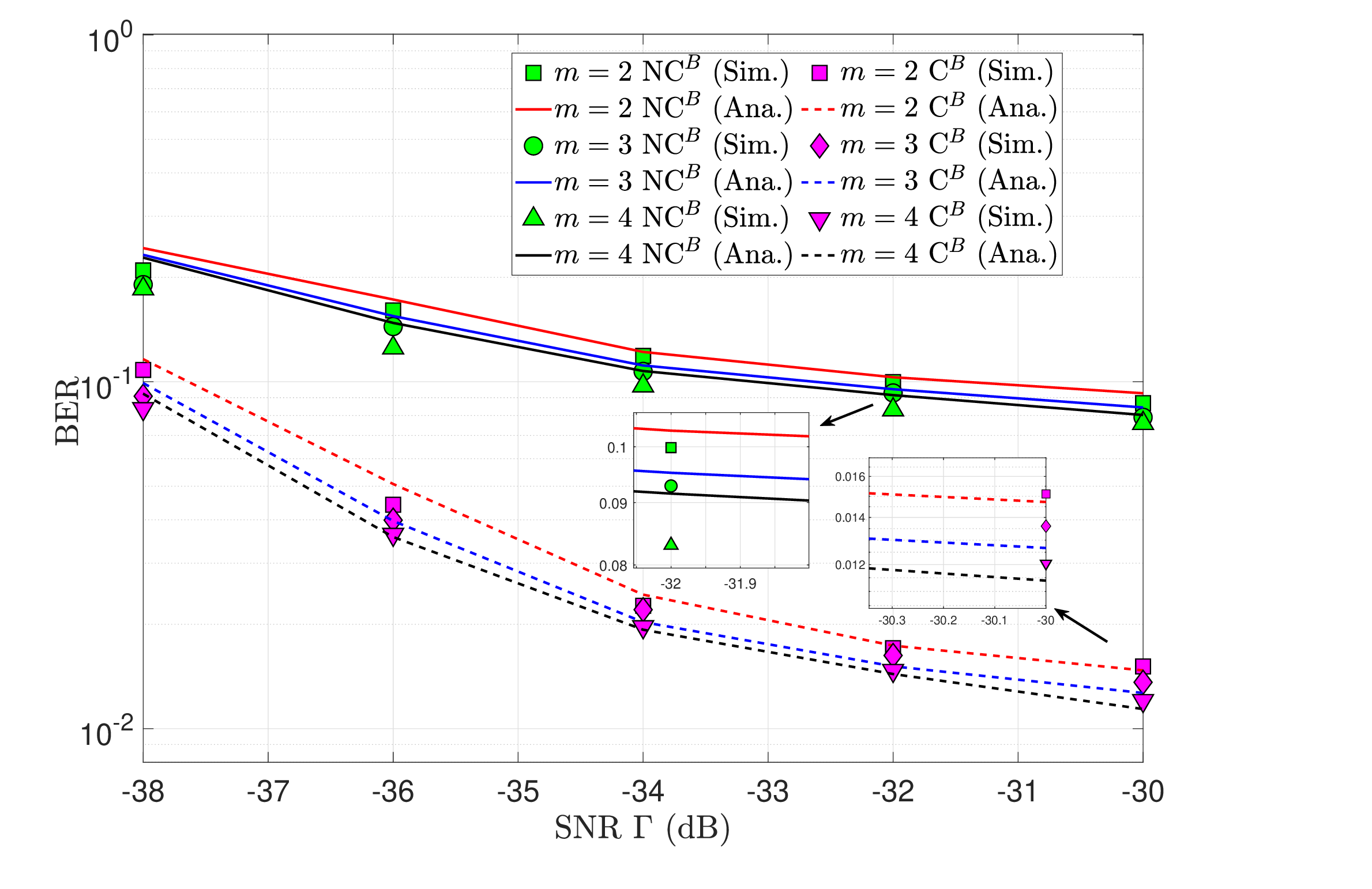}
   \label{BER_case_B}
  }
  \vspace{-0.2cm}
  \caption{Analytical and simulated BER curves for (a) Case A, and (b) Case B, for various values of the shape parameters $m$, where $SF=7$ and $N=20$.} \label{fig:diffm}
 \end{center}
\end{figure}
Fig.~\ref{fig:BER} shows the analytical and simulated BER results of Case A and Case B over Nakagami-$m$ fading channels, where $SF=7,8,9$, $N=25$, and $m = 2$. From the figure, we observe that there is little gap between the analytical curve and the corresponding simulation results, which validates our analytical framework. As shown in Fig.~\ref{fig:BER}, in both Cases A and B for the same SF, coherent detection has a lower BER than non-coherent detection, and increasing the value of $SF$ improves the BER performance. Moreover, one can see that, the BER results for Case A are significantly lower than for Case B. The reason behind this is that, the RIS is used to maximize the target signal only for Case A, while for Case B, the target and interfering signals are maximized by the paired RISs, leading to more interfering power at the gateway. In addition, for Case A, around 2 dB performance gain can be achieved with coherent detection compared to non-coherent detection at the same BER. For Case B, as shown in Fig.~\ref{fig:BER} (b), there is a difference of approximately between the ${10^{ - 1}}$ BER lower bound of coherent detection and non-coherent detection for high SNRs. Therefore, coherent detection is more beneficial under co-SF interference in LoRa systems.

Fig.~\ref{fig:diffREBER} presents the impact of various values of $N$ on the proposed system. It is clearly observed that the BER performances of non-coherent and coherent detections improves as $N$ increases for Case A and Case B. It should be noted that, in conventional RIS-free LoRa systems, a higher SF is usually used to achieve a lower BER and longer transmission distances, but leads to very long transmission times. Therefore, increasing the number of reflective elements to enhance the performance of LoRa systems can enable relatively low SFs to achieve the same robustness as high SFs. Specifically, we compare two scenarios for $SF=7$ and $N=35$, and $SF=8$ and $N=25$. It can be observed that, the BER performance in the scenario of $SF=7$ and $N=35$ outperforms the scenario of $SF=8$ and $N=25$ in Case A, while the BER performance is close to the same in Case B. Thus, increasing the number of reflective elements is an effective measure to handle the trade-off between data rate and transmission distances in RIS-aided LoRa systems.

\begin{figure}[t]
\center
\includegraphics[width=4in,height=2.6in]{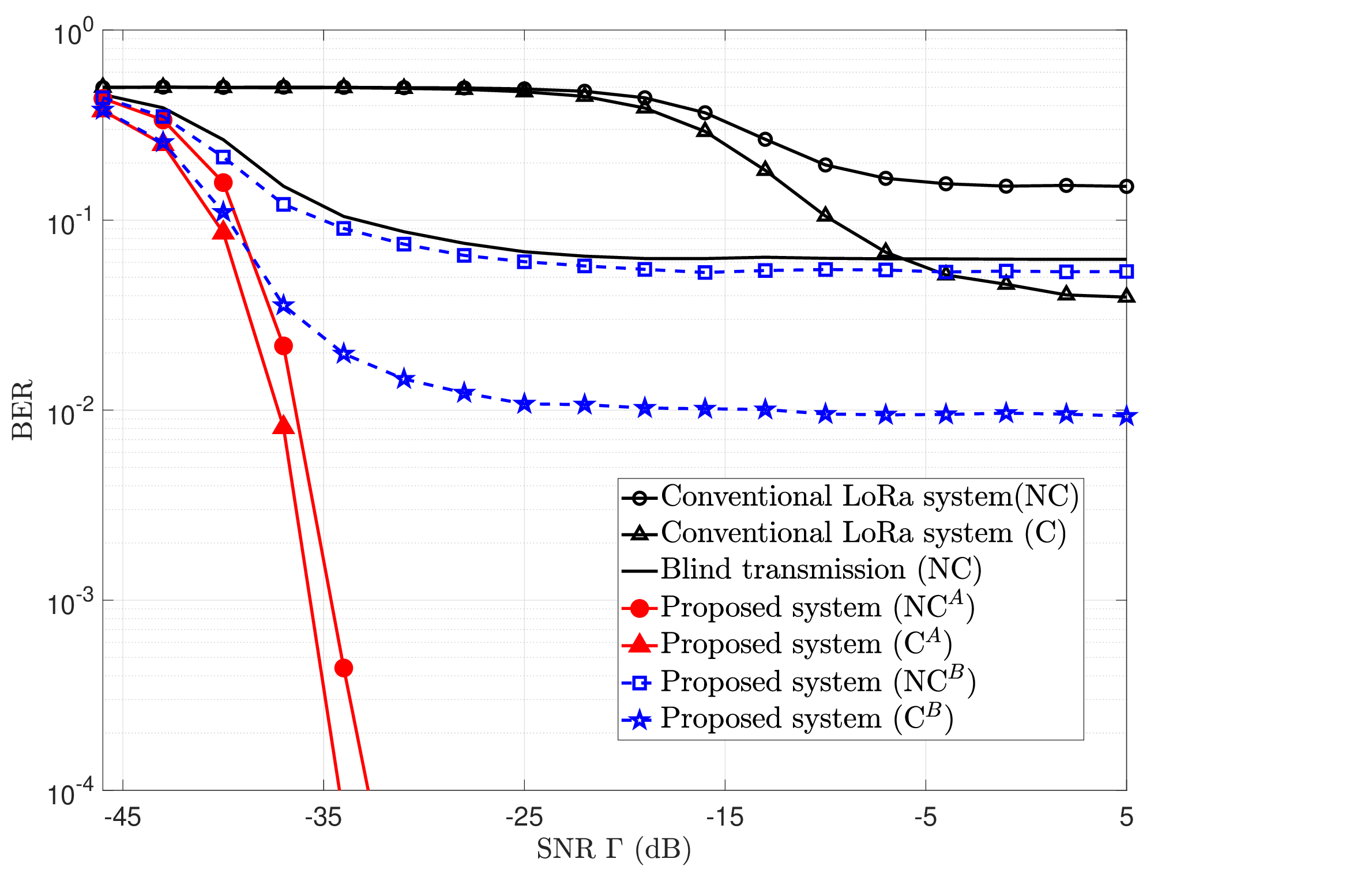}
\caption{{BER comparisons of the conventional LoRa system under interference with non-coherent detection and coherent detection, the RIS-aid LoRa system with interference adopting blind transmission, and the proposed system, where $SF=7$, $m=2$, and $N=25$.}}
\label{fig_vS_conv_blind_compare}
\vspace{-0.45cm}
\end{figure} 	
Fig.~\ref{fig:diffm} presents the BER performance of Case A and Case B for various values of $m$, where $SF=7$ and $N=20$. It is observed that, at low SNRs, the differences between the BER in the case of $m=2,3,4$ are small both for Case A and Case B. However, at high SNRs, as $m$ increases (i.e., communication links have a better LOS propagation environment), the BER performance of both for Case A and Case B is remarkably enhanced. In addition, it is observed that the gain achieved by increasing $m$ is significantly different for the two cases. For example, at $\Gamma  =  - 32$ dB, the BER for Case A with non-coherent detection is $4.8 \times {10^{ - 5}}$ and $1.1 \times {10^{ - 5}}$ for $m = 2$ and $m=3$, respectively. In other words, the BER is $4.5$ times lower when $m$ increases from $2$ to $3$. At $\Gamma  =  - 32$ dB, the BER for Case B with non-coherent detection is $1.0 \times {10^{ - 1}}$ and $9.5 \times {10^{ - 2}}$ for $m = 2$ and $m=3$, respectively. 
Similar gain results as $m$ increases appear in coherent detection both for Case A and Case B. Therefore, the different fading parameter $m$ has a significant impact on the system performance, especially in Case A.

Fig.~\ref{fig_vS_conv_blind_compare} shows the BER performance of the conventional LoRa system under interference with non-coherent and coherent detections, the RIS-aided LoRa system with interference adopting blind transmission, and the proposed system, where $SF = 7$, $m=2$, and $N = 25$.
Since blind transmission in the RIS-aided LoRa system does not require CSI for all communication links (i.e., ${\omega _{T\ell }}$ and ${\omega _{I\ell }}$ both follow uniform distributions in $[0,2\pi)$, $\forall \ell $), we only consider the non-coherent detection.
It is noted that for both Case A and Case B, the signal characteristics at the gateway for blind transmission are the same. As can be seen from this figure, introducing RISs into the LoRa system with interference causes a rapid drop in the BER compared to the conventional RIS-free LoRa system with interference.
In other words, less transmitted power is needed to achieve the performance of a RIS-free LoRa system.
Although the RIS controller consumes energy to control the RIS, this energy consumptions is far less than the transmitted power.
In addition, it is observed that the proposed system outperforms the blind transmission.
Thus, using the channel estimation techniques to perform phase shift configuration of the RIS is an effective measure to improve the performance of LoRa in the presence of interference. In addition, we notice that, for $N=25$ in Case B, the performance of non-coherent detection is very close to the BER lower bound of the conventional RIS-free LoRa systems with coherent detection, while the BER of coherent detection is 5 times lower compared to the non-coherent detection.
Therefore, coherent detection in Case B is more suitable for overcoming interference.
\section{Conclusion} \label{sect:conclusion}
In this paper, a RIS-aid LoRa system with co-SF interference over Nakagami-$m$ fading channels with coherent and non-coherent detections was considered. The proposed system utilizes advanced RIS technology to improve the performance of LoRa systems under interference by improving the wireless environment.
Closed-form BER expressions were derived and corroborated through simulation results. Numerical results proved that, for the proposed system, coherent detection shows a performance advantage against co-SF interference compared to non-coherent detection. Moreover, the BER performance of the proposed system improves as the number of reflecting elements increases, which is a suitable approach to handle the trade-off data rate and coverage range in the proposed system. In addition, the fading parameter $m$ has a significant impact on the BER performance, especially in the case of the target user and the co-SF interfering user adopting the same RIS to deliver information at high SNRs. Finally, we compared the proposed system to the conventional LoRa system with interference and the RIS-aided LoRa system adopting blind transmission. Results show that the proposed system not only saves power but also offer better BER performance in comparison with the above systems.
Thanks to the aforementioned advantages, the proposed system can be considered as a potential solution to reduce the interference and combat fading effects in future LoRa networks.
\appendices
\section{} \label{Hdistribution}
Since $\left| {{h_{Td}}} \right|\sim{\rm{Nakagami}}\left( {{m_1},1} \right)$, the $\iota$-th moment of $\left| {{h_{Td}}} \right|$ can be obtained as
\begin{align}
\label{HTd_lthmoment}
{\mu _{\left| {{h_{Td}}} \right|}}\left( \iota  \right) = \frac{{\gamma \left( {{m_1} + {\iota  \mathord{\left/
 {\vphantom {\iota  2}} \right.
 \kern-\nulldelimiterspace} 2}} \right)}}{{\gamma \left( {{m_1}} \right)}}m_1^{^{ - \frac{\iota }{2}}}.
\end{align}

For simplicity, let ${\zeta _{T\ell }} = \left| {{h_{T\ell }}} \right|\left| {{g_{T\ell }}} \right|$, and ${J_T} = \sum\nolimits_{\ell  = 1}^N {{\zeta _{T\ell }}}$. We focus on the distribution of ${\zeta _{T\ell }}$ firstly. Using $f_{X Y}(z)=\int_{0}^{\infty} \frac{1}{x} f_{Y}\left(\frac{z}{x}\right) f_{X}(x) d x$ for the product of two RVs, the PDF of the ${\zeta _{Tl}}$ can be expressed as ${f_{{\zeta _{Tl}}}}(z) = \int_0^\infty  {\frac{1}{x}} {f_{\left| {{h_{T\ell }}} \right|}}\left( {\frac{z}{x}} \right){f_{\left| {{g_{T\ell }}} \right|}}(x)dx$. Since $\left| {{h_{T\ell }}} \right|\sim{\rm{Nakagami}}\left( {{m_h}_{_{T }},1} \right)$ and $\left| {{g_{T\ell }}} \right|\sim{\rm{Nakagami}}\left( {{m_g}_{_{T }},1} \right)$, the PDF of ${\zeta _{T\ell}}$ can be computed as
\begin{align}
\label{one_joint_REPDF}
{f_{{\zeta _{T\ell}}}}(x) &= \frac{{4m_{{h_T}}^{{m_h}_{_T}}m_{{g_T}}^{{m_g}_{_T}}}}{{\gamma \left( {{m_{{h_T}}}} \right)\gamma \left( {{m_{{g_T}}}} \right)}}{x^{2{m_{{h_T}}} - 1}}\nonumber\\
&\times\int_0^\infty  {{x^{2{m_{{g_T}}} - 2{m_{{h_T}}} - 1}}} \exp \left( { - {m_{{h_T}}}\frac{{{x^2}}}{{{t^2}}} - {m_{{g_T}}}{t^2}} \right)dt.
\end{align}
Using \cite[eq. 3.478.4]{zwillinger2007table}, (\ref{one_joint_REPDF}) can be calculated as
\begin{align}
\label{GeneralizeKG_distribution}
{f_{{\zeta _{T\ell }}}}(x) = \frac{{4\Delta _T^{{m_{{h_T}}} + {m_{{g_T}}}}}}{{\gamma \left( {{m_{{h_T}}}} \right)\gamma \left( {{m_{{g_T}}}} \right)}}{x^{{m_{{h_T}}} + {m_{{g_T}}} - 1}}{G_{{m_{{h_T}}} - {m_{{g_T}}}}}\left( {2{\Delta _T}x} \right),
\end{align}
where ${\Delta _T} = \sqrt {{m_{{h_T}}}{m_{{g_T}}}}$.
Thus, ${\zeta _{T\ell}}$ is a Generalized-$K$ RV with the parameters ${{m_h}_{_T}}$ and ${{m_g}_{_T}}$. Starting from (\ref{GeneralizeKG_distribution}) and after mathematical expectation operations, the $\iota $-th moment of ${\zeta _{T\ell }}$ can be expressed as
${\mu _{{\zeta _T}_\ell }}\left( \iota  \right) = \Delta _T^{ - \iota }\frac{{\gamma \left( {{m_{{h_T}}} + {\iota  \mathord{\left/
 {\vphantom {\iota  2}} \right.
 \kern-\nulldelimiterspace} 2}} \right)\gamma \left( {{m_{{g_T}}} + {\iota  \mathord{\left/
 {\vphantom {\iota  2}} \right.
 \kern-\nulldelimiterspace} 2}} \right)}}{{\gamma \left( {{m_{{h_T}}}} \right)\gamma \left( {{m_{{g_T}}}} \right)}}$.
Thus, the first and second moments of ${J_T}$ can be, respectively, as
\begin{align}
\label{NRE_1th}
{\mu _{{J_T}}}\left( 1 \right) = N \times {\mu _{{\zeta _{T\ell }}}}\left( 1 \right),
\end{align}
\begin{align}
\label{NRE_2th}
{\mu _{{J_T}}}\left( 2 \right) = N \times {\mu _{{\zeta _{T\ell }}}}\left( 2 \right) + N \times \left( {N - 1} \right) \times {\left[ {{\mu _{{\zeta _{T\ell }}}}\left( 1 \right)} \right]^2}.
\end{align}
Based on (\ref{HTd_lthmoment}), (\ref{NRE_1th}), and (\ref{NRE_2th}), the first and second moments of $\left| H \right|$ can be, respectively, as
\begin{align}
\label{H_1th}
{\mu _{\left| H \right|}}\left( 1 \right) = {\mu _{\left| {{h_{Td}}} \right|}}\left( 1 \right) + {\mu _{{J_T}}}\left( 1 \right),
\end{align}
\begin{align}
\label{H_2th}
{\mu _{\left| H \right|}}\left( 2 \right) = {\mu _{\left| {{h_{Td}}} \right|}}\left( 2 \right) + {\mu _{{J_T}}}\left( 2 \right) + 2{\mu _{\left| {{h_{Td}}} \right|}}\left( 1 \right){\mu _{{J_T}}}\left( 1 \right).
\end{align}

\section{} \label{Derivation_HA}
Let $G = \sum\nolimits_{\ell  = 1}^N {{h_{I\ell }}\exp (j{\omega _{T\ell }}){g_{T\ell }}} $, then the distribution of $G$ can be approximated as a complex Gaussian distribution by using the CLT on the sum of complex unit-power independent and identically distributed (i.i.d.) RVs, i.e., $G \sim \mathcal{CN} \left( {0,N} \right)$. The distribution of $\left| G \right|$ is given by
$\left| G \right| \sim {\rm{Nakagami}}\left( {1,N} \right)$ since a Nakagami RV is equivalent to a Rayleigh RV for $m=1$. The $\iota$-th moment of $\left| {{h_{Id}}} \right|$ and $\left| G \right|$ can be respectively obtained as ${\mu _{\left| {{h_{Id}}} \right|}}\left( \iota  \right) = \frac{{\gamma \left( {{m_2} + {\iota  \mathord{\left/
 {\vphantom {\iota  2}} \right.
 \kern-\nulldelimiterspace} 2}} \right)}}{{\gamma \left( {{m_2}} \right)}}m_2^{{\iota  \mathord{\left/
 {\vphantom {\iota  2}} \right.
 \kern-\nulldelimiterspace} 2}}$ and
 ${\mu _{\left| G \right|}}\left( \iota  \right) = \gamma \left( {1 + \frac{\iota }{2}} \right){N^{{\iota  \mathord{\left/
 {\vphantom {\iota  2}} \right.
 \kern-\nulldelimiterspace} 2}}}$, respectively. Then, the first and second moments of ${\left| {{H_A}} \right|^2}$ can be respectively obtained as
\begin{align}
\label{HA_1th}
{\mu _{{{\left| {{H_A}} \right|}^2}}}\left( 1 \right) = {\mu _{{{\left| {{h_{Id}}} \right|}^2}}}\left( 2 \right) + {\mu _{{{\left| G \right|}^2}}}\left( 2 \right),
\end{align}
\begin{align}
\label{HA_2th}
{\mu _{{{\left| {{H_A}} \right|}^2}}}\left( 2 \right) &= {\rm{E}}\left\{ {{{\left| {{h_{Id}}} \right|}^4} \!+\! {{\left| G \right|}^4}\! + 2\left| G \right|^2{{\left| {{h_{Id}}} \right|}^2} \!+ \!4{{\left| {{h_{Id}}} \right|}^2}\Re \left( {h_{Id}^*G} \right)} \right. \nonumber \\
&\left. { + 4{{\left| G \right|}^2}\Re \left( {h_{Id}^*G} \right) + 4{{\left[ {\Re \left( {h_{Id}^*G} \right)} \right]}^2}} \right\}.
\end{align}
 It is assumed that the real and imaginary parts of ${{h_{Id}}}$ and $G$ have independence, zero-mean and numerical character symmetry characteristics, resulting in ${\rm{E}}\left[ {\Re \left( {h_{Id}^*G} \right)} \right] = 0$, and ${\rm{E}}\left\{ {{{\left[ {\Re \left( {h_{Id}^*G} \right)} \right]}^2}} \right\} = \frac{{{\rm{E}}\left( {{{\left| {{h_{Id}}} \right|}^2}} \right){\rm{E}}\left( {{{\left| G \right|}^2}} \right)}}{2}$. In addition, due to the independence property between ${\left| G \right|}$ and ${{{\left| {{h_{Id}}} \right|}^2}}$ , (\ref{HA_2th}) can be reexpressed as
\begin{align}
\label{HA_2th1}
{\mu _{{{\left| {{H_A}} \right|}^2}}}\left( 2 \right) = {\mu _{\left| {{h_{Id}}} \right|}}\left( 4 \right){\rm{ + }}{\mu _{\left| G \right|}}\left( 4 \right) + 4{\mu _{\left| {{h_{Id}}} \right|}}\left( 2 \right){\mu _{\left| G \right|}}\left( 2 \right).
\end{align}

Based on (\ref{HA_1th}) and (\ref{HA_2th1}), the estimator of ${\omega _A}$ and ${\lambda _A}$ can be, respectively, obtained as
\begin{align}
\label{omigaA}
{\omega _A} & =  {\frac{{{{\left[ {{\mu _{{{\left| {{H_A}} \right|}^2}}}\left( 1 \right)} \right]}^2}}}{{{\mu _{{{\left| {H{_A}} \right|}^2}}}\left( 2 \right) - {\mu _{{{\left| {{H_A}} \right|}^2}}}\left( 1 \right)}}} ,
\end{align}
\begin{align}
\label{lamdaA}
{\lambda _A} =  {\frac{{{\mu _{{{\left| {H{_A}} \right|}^2}}}\left( 1 \right)}}{{{\mu _{{{\left| {H{_A}} \right|}^2}}}\left( 2 \right) - {\mu _{{{\left| {H{_A}} \right|}^2}}}\left( 1 \right)}}} .
\end{align}
\section{}\label{closed-form_Gauss_app}
Here, some closed-form SER expressions are derived.
According to (\ref{Interfererence_closefrom_caseA_Ncoherent}), we have
\begin{align}
\label{APPENDIX C_caseA+Ncoherent}
P_{{A_{{\rm{NC}}}}}^{{\rm{I}}|I,\tau } &\approx \frac{{\lambda _H^{{\omega _H}}}}{{2\gamma \left( {{\omega _H}} \right)}}\frac{{\lambda _A^{{\omega _A}}}}{{\gamma \left( {{\omega _A}} \right)}}\int_0^\infty  {\int_0^\infty  {Q\left( {\frac{{\sqrt {x{E_s}}  - \sqrt {y{E_I}} {\chi _I}}}{{\sqrt {{N_0}} }}} \right)} }\nonumber \\&\times{\left( {\sqrt x } \right)^{{\omega _H} - 2}}{y^{{\omega _H} - 1}}\exp \left( { - {\lambda _H}\sqrt x  - {\lambda _A}y} \right)dxdy.
\end{align}
Due to the complexity of the double integral, we introduce the Gauss-Hermite quadrature method \cite[Table 25.10]{zwillinger2007table}, i.e.,
\begin{align}
\int_0^\infty  {p\left( x \right)} dx = \sum\limits_{\upsilon  = 1}^V {{\psi _\upsilon }} p\left( {{\varpi _\upsilon }} \right)\exp \left( {\varpi _\upsilon ^2} \right) + {O_V} ,
\end{align}
where $V$ is the number of sample points for approximation, ${{x_\upsilon }}$ is the $\upsilon$-th root of the Hermite polynomial ${H_V}\left( x \right)\left( {\upsilon  = 1,2,...,V} \right)$, ${{\psi _\upsilon }}$ is the $\upsilon$-th  associated weight obtained from $\sqrt \pi  \frac{{{2^{V - 1}}V!}}{{{V^2}H_{W - 1}^2\left( {{\varpi _\upsilon }} \right)}}$, and ${O_V}$ is the remainder term which tends to zero when $V$ is large enough. It is noted that, in the following analysis, we take a large value for $V$ to obtain more accurate outcomes, resulting in ${O_V} \approx 0$. Then, we focus on the integral variable $x$. Let $F = \sqrt {\frac{{y{E_I}}}{{{N_0}}}} {\chi _I}$. Variable substitution $\sqrt x  = \exp \left( \alpha  \right)$ is used to obtain the new limits of the integral from $ - \infty $ to $ \infty$. After some mathematical manipulations, the first integral with respect to $x$ can be calculated as
\begin{align}
\label{APPENDIX C_first_integral}
\int_0^\infty  &{Q\left( {\sqrt {\frac{{x{E_I}}}{{{N_0}}}}  - F} \right)} {\left( {\sqrt x } \right)^{{\omega _H} - 2}}\exp \left( { - {\lambda _H}\sqrt x } \right)dx \nonumber \\
&\approx 2\sum\limits_{{\upsilon _1} = 1}^{{V_1}} {{\psi _\upsilon }_{_1}} Q\left[ {\sqrt {\Gamma K} \exp \left( {{\alpha _\upsilon }_{_1}} \right) - F} \right]\nonumber \\
&\times\exp \left[ {\alpha _{{\upsilon _1}}^2 - {\lambda _H}\exp \left( {{\alpha _\upsilon }_{_1}} \right) + {\alpha _\upsilon }_{_1}{\omega _H}} \right].
\end{align}
Substituting $F$ into (\ref{APPENDIX C_first_integral}), and using variable substitution $y = \exp \left( \beta  \right)$, relying on the Gauss-Hermite quadrature method again, and performing some mathematical manipulations, (\ref{APPENDIX C_caseA+Ncoherent}) can be obtained as
\begin{align}
\label{IANC_first_Gaussian}
P_{{A_{{\rm{NC}}}}}^{{\rm{I}}|I,\tau } &\approx \frac{{\lambda _H^{{\omega _H}}}}{{\gamma \left( {{\omega _H}} \right)}}\frac{{\lambda _A^{{\omega _A}}}}{{\gamma \left( {{\omega _A}} \right)}}\sum\limits_{{\upsilon _2} = 1}^{{V_2}} {\sum\limits_{{\upsilon _1} = 1}^{{V_1}} {{\psi _\upsilon }_{_2}{\psi _\upsilon }_{_1} \times } } \
\nonumber\\&\times Q\left[ {\sqrt {\Gamma K} \exp \left( {{\alpha _\upsilon }_{_1}} \right) - \sqrt {\Gamma K\exp \left( {{\beta _\upsilon }_{_2}} \right)} {\chi _I}} \right]
\nonumber\\& \times \exp \left[ \begin{array}{l}
\alpha _{{\upsilon _1}}^2 + {\omega _H}{\alpha _\upsilon }_{_1} - {\lambda _H}\exp \left( {{\alpha _\upsilon }_{_1}} \right)\\
 + \beta _{{\upsilon _2}}^2 + {\omega _A}{\beta _\upsilon }_2 - {\lambda _A}\exp \left( {{\beta _\upsilon }_{_2}} \right)
\end{array} \right].
\end{align}
Similarly, (\ref{eq:RIS_interferenceA_Cinterfer_closeform_event}) is computed as
\begin{align}
\label{before_staircase}
P_{{A_{\rm{C}}}}^{{\rm{I}}|I,\tau } &\approx \frac{1}{{2\pi }}\frac{{\lambda _H^{{\omega _H}}}}{{\gamma \left( {{\omega _H}} \right)}}\frac{{\lambda _A^{{\omega _A}}}}{{\gamma \left( {{\omega _A}} \right)}}{\sum\limits_{{\upsilon _2} = 1}^{{V_2}} {\sum\limits_{{\upsilon _1} = 1}^{{V_1}} {{\psi _\upsilon }_{_2}{\psi _\upsilon }_{_1}} } }\nonumber \\
&\times\exp \left[
\begin{array}{l}
\alpha _{{\upsilon _1}}^2 + {\omega _H}{\alpha _\upsilon }_{_1} - {\lambda _H}\exp \left( {{\alpha _\upsilon }_{_1}} \right)\\
 + \beta _{{\upsilon _2}}^2 + {\omega _A}{\beta _\upsilon }_{_2} - {\lambda _A}\exp \left( {{\beta _\upsilon }_{_2}} \right)
\end{array} \right]
\nonumber \\
&\times\!\!\int_0^{2\pi }\!\! {Q\!\left[ {\sqrt {\Gamma K} \exp \left( {{\alpha _\upsilon }_{_1}} \right) \!-\! \sqrt {\Gamma K\exp \left( {{\beta _\upsilon }_{_2}} \right)} {\chi _I}\cos \vartheta } \right]} d\vartheta \!.
\end{align}
Utilizing the $M$-staircase approximation \cite{9558795}, (\ref{before_staircase}) is computed as
\begin{align}
\label{before_staircase2}
P_{{A_{\rm{C}}}}^{{\rm{I}}|I,\tau } &\approx \frac{1}{M}\frac{{\lambda _H^{{\omega _H}}}}{{\gamma \left( {{\omega _H}} \right)}}\frac{{\lambda _A^{{\omega _A}}}}{{\gamma \left( {{\omega _A}} \right)}}\sum\limits_{\varsigma = 1}^M {\sum\limits_{{\upsilon _2} = 1}^{{V_2}} {\sum\limits_{{\upsilon _1} = 1}^{{V_1}} {{\psi _\upsilon }_{_2}{\psi _\upsilon }_{_1}} }}\nonumber\\ &\times
 Q\left[ {\sqrt {\Gamma K} \exp \left( {{\alpha _\upsilon }_{_1}} \right) - \sqrt {\Gamma K\exp \left( {{\beta _\upsilon }_{_2}} \right)} \cos \frac{{2\pi \varsigma}}{M}{\chi _I}} \right]\nonumber\\ &\times\exp \left[ \begin{array}{l}
\alpha _{{\upsilon _1}}^2 + {\omega _H}{\alpha _\upsilon }_{_1} - {\lambda _H}\exp \left( {{\alpha _\upsilon }_{_1}} \right)\\
 + \beta _{{\upsilon _2}}^2 + {\omega _A}{\beta _\upsilon }_{_2} - {\lambda _A}\exp \left( {{\beta _\upsilon }_{_2}} \right)
\end{array} \right]  .
\end{align}
To achieve more accurate results, we set $V_1=V_2=70$, and $M=20$.
Similar to the derivation of (\ref{IANC_first_Gaussian}), (\ref{CaseB_NC_integral}) can be calculated as
\begin{align}
\label{APPENDIX_caseB_NC_closeform}
P_{{B_{{\rm{NC}}}}}^{{\rm{I}}|I,\tau }& \approx \frac{{\lambda _H^{2{\omega _H}}}}{{{{\left[ {\gamma \left( {{\omega _H}} \right)} \right]}^2}}}\sum\limits_{{\upsilon _2} = 1}^{{V_2}} {\sum\limits_{{\upsilon _1} = 1}^{{V_1}} {{\psi _\upsilon }_{_2}{\psi _\upsilon }_{_1}}}  \\ &\times Q\left( {\sqrt {\Gamma K} \exp \left( {{\alpha _\upsilon }_{_1}} \right) - \sqrt {\Gamma K} \exp \left( {{\beta _\upsilon }_{_2}} \right){\chi _I}} \right) \nonumber\\& \times \exp \left\{ \begin{array}{l}
\alpha _{{\upsilon _1}}^2 + \beta _{{\upsilon _2}}^2 + {\omega _H}\left( {{\alpha _\upsilon }_1 + {\beta _\upsilon }_{_2}} \right)\\
 - {\lambda _H}\left[ {\exp \left( {{\alpha _\upsilon }_1} \right) + \exp \left( {{\beta _\upsilon }_2} \right)} \right]
\end{array} \right\}.
\end{align}

Moreover, following the same approach as for (\ref{before_staircase2}), (\ref{caseB_coherent_integral_inter}) can be computed as
\begin{align}
P_{{B_{\rm{C}}}}^{{\rm{I}}|I,\tau } &\approx \frac{1}{M}\frac{{\lambda _H^{2{\omega _H}}}}{{{{\left[ {\gamma \left( {{\omega _H}} \right)} \right]}^2}}}\sum\limits_{\varsigma  = 1}^M {\sum\limits_{{\upsilon _2} = 1}^{{V_2}} {\sum\limits_{{\upsilon _1} = 1}^{{V_1}} {{\psi _\upsilon }_{_2}{\psi _\upsilon }_{_1}} }}\nonumber\\
&\times Q\left[ {\sqrt {\Gamma K} \exp \left( {{\alpha _\upsilon }_{_1}} \right) - \sqrt {\Gamma K} \exp \left( {{\beta _\upsilon }_{_2}} \right)\cos \frac{{2\pi \varsigma }}{M}{\chi _I}} \right]\nonumber\\
& \times \exp \left\{ \begin{array}{l}
\alpha _{{\upsilon _1}}^2 + \beta _{{\upsilon _2}}^2 + {\omega _H}\left( {{\alpha _\upsilon }_1 + {\beta _\upsilon }_{_2}} \right)\\
 - {\lambda _H}\left[ {\exp \left( {{\alpha _\upsilon }_1} \right) + \exp \left( {{\beta _\upsilon }_2} \right)} \right]
\end{array} \right\}.
\end{align}
\vspace{-0.45cm}


\end{document}